\documentclass[a4paper,12pt]{article}

\usepackage{cite}

\usepackage{fullpage}
\usepackage[font=footnotesize,labelfont=bf,margin=1cm]{caption}
\usepackage{amsfonts}
\usepackage{amsmath}
\usepackage{amssymb}

\numberwithin{equation}{section}
\allowdisplaybreaks[4]

\newcommand{\qandq}{\qquad\mathrm{and}\qquad}
\newcommand{\diag}{\mathrm{diag}}

\newcommand{\dd}{\mathrm{d}}		
\newcommand{\HH}{\mathcal{H}}			
\newcommand{\MM}{\mathcal{M}}			
\newcommand{\EE}{\mathcal{E}}			
\newcommand{\XX}{\mathbb{X}}				

\newcommand{\tx}{\tilde{x}}
\newcommand{\ty}{\tilde{y}}
\newcommand{\tz}{\tilde{z}}
\newcommand{\ttt}{\tilde{t}}
\newcommand{\tC}{\tilde{C}}
\newcommand{\tF}{\tilde{F}}

\newcommand{\tQ}{\tilde{Q}}

\newcommand{\bi}{{\bar{i}}}
\newcommand{\bj}{{\bar{j}}}
\newcommand{\bm}{{\bar{m}}}
\newcommand{\bn}{{\bar{n}}}
\newcommand{\bk}{{\bar{k}}}
\newcommand{\bl}{{\bar{l}}}

\newcommand{\bz}{{\bar{z}}}
\newcommand{\ba}{{\bar{a}}}
\newcommand{\bb}{{\bar{b}}}

\begin{document}

\begin{titlepage}
\vfill

\begin{flushright}
QMUL-PH-14-18
\end{flushright}

\vfill

\begin{center}
   \baselineskip=16pt
   	{\Large \bf Branes are Waves and Monopoles}
   	\vskip 2cm
   	{\sc		David S. Berman\footnote{\tt d.s.berman@qmul.ac.uk} and
	 		Felix J. Rudolph\footnote{\tt f.j.rudolph@qmul.ac.uk}}
	\vskip .6cm
    {\small \it Queen Mary University of London, Centre for Research in String Theory, \\
             School of Physics, Mile End Road, London, E1 4NS, England} \\ 
	\vskip 2cm
\end{center}

\begin{abstract}
In a recent paper it was shown that fundamental strings are null waves in Double Field Theory. Similarly, membranes are waves in exceptional extended geometry. Here the story is continued by showing how various branes are Kaluza-Klein monopoles of these higher dimensional theories. Examining the specific case of the $E_7$ exceptional extended geometry, we see that all branes are both waves and monopoles. Along the way we discuss the $O(d,d)$ transformation of localized brane solutions not associated to an isometry and how true T-duality emerges in Double Field Theory when the background possesses isometries.
\end{abstract}

\vfill

\setcounter{footnote}{0}
\end{titlepage}

\tableofcontents

\section{Introduction}

Double Field Theory and its M-theory generalization, exceptional extended geometry, now have a long history. Following the initial endeavours by Duff \cite{Duff90a,Duff90b}, Tseytlin formulated a version of the string in a doubled space \cite{Tseytlin90,Tseytlin91}. The new geometry to describe a duality covariant version of supergravity was introduced by Siegel \cite{Siegel93a,Siegel93b,Siegel93c}. Hull then constructed the double sigma model in \cite{Hull:2004in} and developed more of the ideas which eventually led to the conception of Double Field Theory (DFT) which was established with the seminal paper \cite{Hull:2009mi} by Hull and Zwiebach. Since DFT allows for dynamics in all the doubled dimensions, it goes beyond the duality covariant formulation of supergravity. DFT was then developed further \cite{Hohm:2010jy,Hohm:2010pp,Hohm:2010xe,Hohm:2011nu} and expanded in various directions by Park and collaborators \cite{Jeon:2010rw,Jeon:2011cn,Jeon:2011vx,Jeon:2011sq} and others \cite{Blair:2013noa,Berman:2013uda,Blair:2014kla,Wu:2013sha,Wu:2013ixa,Ma:2014ala}. Similar developments for the U-duality groups of M-theory can be found in \cite{Hull:2007zu,Pacheco:2008ps,Hillmann:2009ci,Berman:2010is,Coimbra:2011ky,Coimbra:2012af, Berman:2011pe,Berman:2011kg,Berman:2011cg,Berman:2011jh,Berman:2012vc,Park:2013gaj, Cederwall:2013oaa,Cederwall:2013naa,Strickland-Constable:2013xta, Park:2014una}. Of course, from one point of view many of the ideas in DFT and extended geometry were anticipated by the $E_{11}$ programme of West and collaborators; see for example \cite{West:2001as,Englert:2003zs,West:2003fc,Kleinschmidt:2003jf,West:2004kb}. For a review of double field theory and its generalisations one may choose from the following three articles \cite{Berman:2013eva,Hohm:2013bwa,Aldazabal:2013sca}. 

We will now adopt a rather simplistic approach which begins with the question, is there a lift of supergravity to a higher dimensional theory where the p-form potentials are ``geometric'' just as the graviphoton is in conventional Kaluza-Klein theory?

If one only considers the NS-NS sector of ten-dimensional supergravity where there is only the Kalb-Ramond two-form potential, then the answer to this question is Double Field Theory. If one considers eleven-dimensional supergravity with $C_3$ and $C_6$ potentials, then the answer is exceptional extended geometry. The message is that one may view these novel extended theories as lifts of known theories and the so-called ``section condition'' is the Kaluza-Klein reduction constraint. 

In Kaluza-Klein theories the origin of electric charge is from momentum in the KK-direction. The quantization of momentum then results in the quantization of electric charge. The origin of magnetic charges comes from twisting the KK-circle to produce a non-trivial circle bundle with non-vanishing first Chern class. The first Chern class is the magnetic charge. The construction of such a non-trivial solution for traditional Kaluza-Klein theory was first given in \cite{Sorkin:1983ns,Gross:1983hb}.

In 1995 M-theory came into being. A crucial aspect was the lift of Type IIA supergravity to eleven dimensions and the Ramond-Ramond one-form playing the role of the KK-graviphoton. Crucially, it was not just the fields of the Type IIA theory that could be lifted to eleven dimensions but also the charged states. Most notably, the D0-brane was identified with the momentum in the eleventh direction and the D6-brane was identified as the associated KK-monopole\footnote{We are extremely grateful to Paul Townsend who pointed out the importance of the identification of the D6-brane with the monopole in the M-theory context which then inspired this paper.} \cite{Townsend:1995kk}.

Thinking of Double Field Theory as a Kaluza-Klein theory immediately brings forth the idea of describing the fundamental string as a momentum state with the momentum in the novel additional directions. This was the subject of a recent paper \cite{Berkeley:2014nza}. Not only could the string be identified as the null wave solution in DFT, but the effective action of such a solution could be identified with the string action with manifest $O(d,d)$ symmetry \cite{Tseytlin90,Tseytlin91}.

Logically, the remaining task is to identify the monopole-like solutions in DFT. That is, what do the DFT equivalents of KK-monopole solutions correspond to? It should not be a surprise to the reader that this is the NS5-brane since it is the magnetic dual to the fundamental string. The NS5-brane is also in the same $O(d,d)$ orbit as the KK-monopole solution in supergravity and so is a natural candidate. Note that technically there is something much more non-trivial about having a mono-pole-type solution of DFT than the null wave solution. The null wave has a trivial dilaton whereas the fivebrane does not which leads to more complicated equations of motion. 

It is a pleasure to note that the fivebrane/monopole in DFT has been previously studied in some very inspiring articles using gauged linear sigma model techniques, originally by Jensen \cite{Jensen:2011jna} and later in various detailed works by Kimura \cite{Kimura:2013fda,Kimura:2013zva,Kimura:2013khz,Kimura:2014upa}. 

When one talks of the DFT monopole solution, one is describing a solution of DFT with a particular monopole-like ansatz for the generalized metric. In the case where there is an isometry so that one has two T-duality related solutions of supergravity -- the NS5-brane and the KK-monopole -- this is an embedding of those solutions in DFT. When there is no isometry (though topologically there is  a circle) then we have localized solutions in DFT which require the existence of solutions with no supergravity description. Thus these localized solutions cannot be purely thought of as an embedding in DFT since they correspond to solutions with no ordinary spacetime interpretation.  

Having identified the NS5-brane as the KK-monopole in DFT, one may then repeat this trick with the exceptional extended geometry and describe the M-theory fivebrane as a KK-monopole in the exceptional theory. This should not be so much of a surprise since the M-theory lift of the NS5-brane is the M5-brane. However we can also consider the membrane as a KK-monopole of the exceptional theory which is perhaps some what more surprising. Finally, we can also have the fivebrane as a null wave. Thus in the exceptional geometry case, the membrane and fivebrane solutions of eleven-dimensional supergravity may be identified as either a wave- or a monopole-like solution of the extended theory. 

On further reflection, this had to be the case since the whole point of the exceptional extended geometry is to have U-duality manifest symmetry of the theory. S-duality is clearly a part of the U-duality group. S-duality swaps ``electric'' and ``magnetic'' solutions which in terms of geometry means exchanging null wave solutions with monopole like solutions. This is a non-trivial duality since it relates solutions with different topology. 

The story of this paper is similar to what happens in the six-dimensional $(0,2)$ theory associated with the M-theory fivebrane. The $(0,2)$ theory is self-dual in six dimensions and under dimensional reduction on a torus this self-duality results in the hidden duality symmetry of the lower dimensional theory, such as the S-duality in four-dimensional $\mathcal{N}=4$ Super-Yang-Mills \cite{Verlinde:1995mz,Witten:1995gf}. The relevant solution of the six-dimensional theory is the self-dual string. It is only how one identifies the wrapped self-dual string with states in the four-dimensional theory that causes the emergence of the hidden duality symmetry.

Just like the $(0,2)$ theory, the exceptional extended geometry is describing a theory where the duality group is a manifest symmetry. As such it is only through the reduction to the lower dimensional theory that one actually produces a hidden duality. What is novel is that this is a gravitational theory as opposed to the field theory examples that have been studied so far and the duality group is beyond that of the $SL(2)$ corresponding to large diffeomorphism of the torus. Yet the principle is the same. In general we expect all solutions related under U-duality to be a single solution in the extended geometry.

Let us start by describing the monopole in DFT and using this to extract the NS5-brane. We will then show how the M-theory fivebrane may be described in the exceptional extended geometry associated to $E_7$ first as a null wave and then as a monopole solution. We will then also show how the membrane can be produced as both wave and monopole solutions. Finally we comment further on the implications.

\section{The Monopole in DFT}
\label{sec:DFT}
In what follows it will be useful to introduce coordinates $(x^\mu,\tx_\mu)$ for Double Field Theory. We will call the coordinates associated to our usual notion of spacetime $x^\mu$ and the {\it{winding}} or {\it{dual}} coordinates $\tx_\mu$. It is the presence of the $O(d,d)$ structure $\eta$ that allows this split into $(x^\mu,\tx_\mu)$ coordinates since $\eta$ produces a natural pairing between coordinates. (For the reader familiar with the symplectic geometry of classical mechanics, $\eta$ is very much like a symplectic form and may be used to define a polarization which is essentially what one does when applying the section condition or equivalently picking a duality frame.) The action and equations of motion of DFT are concisely written in Appendix \ref{sec:appDFT} for easy referral.

In \cite{Berkeley:2014nza} a null wave in the doubled space of DFT was shown to reduce to a pp-wave or a fundamental string when viewed from the ordinary supergravity point of view. The interpretation of the solution in terms of the normal supergravity theory associated to the reduction of DFT was determined by the direction the null wave was travelling in. If the DFT solution carries momentum in a spacetime direction $x$ it reduces to a wave. But if it carries momentum in a dual (winding) direction $\tilde{x}$ it gives the string whose mass and charge are determined by the momentum in that dual direction. 

Instead of the wave we will now consider the Kaluza-Klein monopole solution also known as the Sorkin-Gross-Perry monopole \cite{Sorkin:1983ns,Gross:1983hb}
\begin{equation}
\begin{aligned}
\dd s^2 &= H^{-1}\left[\dd z + A_i\dd y^i\right]^2 + H\delta_{ij}\dd y^i\dd y^j \\
H &= 1 + \frac{h}{|\vec{y}_{(3)}|} \, ,  \qquad
\partial_{[i}A_{j]} = \frac{1}{2}{\epsilon_{ij}}^k\partial_k H 
\end{aligned}
\end{equation} 
where $H$ is a harmonic function and $A_i$ a vector potential with $i=1,2,3$. If this solution is supplemented by some trivial world volume directions, it can be turned into something known as a KK-brane, the KK-monopole being a KK0-brane. The low energy limit of M-theory is eleven-dimensional supergravity. Thus, to embed the monopole solution (which is four-dimensional) requires adding seven trivial dimensions (one of which is timelike) which would then produce a KK6-brane solution as follows
\begin{equation}
\dd s^2 = -\dd t^2 + \dd\vec{x}_{(6)}^{\, 2} 
				+ H^{-1}\left[\dd z + A_i\dd y^i\right]^2 + H\dd\vec{y}_{(3)}^{\, 2}
\end{equation}
where $H$ and $A_i$ are the same as above. (From the point of view of Type IIA supergravity, which is the theory that emerges upon Kaluza-Klein reduction in the $z$ direction, this is the Type IIA D6-brane.)  All of this is part of the usual supergravity story relating solutions of eleven-dimensional supergravity to those of the Type IIA theory \cite{Townsend:1995kk}.

Now let us consider a monopole-type solution in Double Field Theory which we call the {\it{DFT monopole}}. Appendix \ref{sec:appDFT} shows that the following is a solution and satisfies the DFT equations of motion. The solution is described by the generalized metric $\HH_{MN}$ given below. It is an open question if the generalized metric is an actual metric tensor on the doubled space or something different, in which case the term ``metric'' is a misnomer. For the purpose of this paper it is sufficient that the generalized metric transforms under generalized diffeomorphisms (generated by the generalized Lie derivative) and for a given solution satisfies the DFT equations of motion. Nevertheless, for convenience we will encode the matrix $\HH_{MN}$ in terms of a ``line element'' $\dd s^2 = \HH_{MN}\dd X^M \dd X^N$ which provides a concise way of presenting the components of $\HH_{MN}$. It is not necessary to us that this line element defines an actual metric tensor in the doubled space{\footnote{We thank Chris Hull for emphasizing this issue to us.}}. With this caveat in mind, we write the monopole solution as follows
\begin{equation}
\begin{aligned}
\dd s^2 &= \HH_{MN}\dd X^M \dd X^N \\
	&= H(1+H^{-2}A^2) \dd z^2 + H^{-1} \dd \tz^2 
			+ 2H^{-1} A_i[\dd y^i \dd \tz - \delta^{ij} \dd \ty_j \dd z ]\\
	&\quad + H(\delta_{ij}+H^{-2}A_iA_j) \dd y^i \dd y^j 
			+ H^{-1} \delta^{ij} \dd \ty_i \dd \ty_j \\
	&\quad +\eta_{ab}\dd x^a \dd x^b + \eta^{ab}\dd \tx_a \dd \tx_b 
\end{aligned}
\label{eq:DFTmonopole}
\end{equation}
and the rescaled dilaton of DFT (defined as $e^{-2d}=g^{1/2}e^{-2\phi}$) 
\begin{equation}
e^{-2d} = He^{-2\phi_0} 
\end{equation}
where $\phi_0$ is a constant. The generalized coordinates with $M=1,\dots,20$ are
\begin{equation}
X^M = (z,\tz,y^i,\ty_i,x^a,\tx_a)
\end{equation}
where $i=1,2,3$ and $a=1,\dots,6$. The last line in the line element uses the Minkowski metric $\eta_{ab}$, i.e. $x^1=t$ and $\tx_1=\ttt$ are timelike, our signature is mostly plus. 

Here $H$ is a harmonic function of the $y^i$ only; it is annihilated (up to delta function sources) by the Laplacian in the $y$-directions and is given by
\begin{equation}
H(r) = 1 + \frac{h}{r}\, , \qquad r^2 = \delta_{ij}y^iy^j
\end{equation} 
with $h$ an arbitrary constant that is related to the magnetic charge. The vector $A_i$ also obeys the Laplace equation, is divergence-free and its curl is given by the gradient of $H$
\begin{equation}
\vec{\nabla}\times\vec{A} = \vec{\nabla} H 
\qquad\mathrm{or}\qquad
\partial_{[i}A_{j]} = \frac{1}{2}{\epsilon_{ij}}^k\partial_k H \, .
\label{eq:AH}
\end{equation}

This doubled solution is to be interpreted as a KK-brane of DFT. It can be rewritten to extract the spacetime metric $g_{\mu\nu}$ and the Kalb-Ramond two-form $B_{\mu\nu}$ in ordinary spacetime with coordinates $x^\mu=(z,y^i,x^a)$. We will show explicitly that the ``reduced'' solution is in fact an infinite periodic array of NS5-branes smeared along the $z$ direction.

One can also show that if $\tz$ is treated as a normal coordinate and $z$ as a dual coordinate the reduced solution is the string theory monopole introduced above. This means the (smeared) NS5-brane is the same as a KK-monopole with the KK-circle in a dual (winding) direction.

One might be concerned about the presence of $A_i$ in the generalized metric since for the monopole picture to make sense, $A_i$ must transform as a one-form gauge field. (Below we show how this one-form is a component of the two-form $B_{\mu \nu}$). Crucially, the generalized metric transforms under the so-called generalized Lie derivative. When the generating double vector field of the generalized Lie derivative points in the dual space directions it generates the gauge transformations of the B-field. When we have an additional isometry, the $z$ direction of this solution, then this generalized Lie derivative generates the correct gauge transformations of a one-form field $A_i$. (This requires the gauge parameters to also be independent of $z$).

\subsection{Rewriting the Solution}
We will now use the form of the doubled metric $\HH_{MN}$ in terms of $g_{\mu\nu}$ and $B_{\mu\nu}$ to rewrite the solution \eqref{eq:DFTmonopole} in terms of ten-dimensional non-doubled quantities. This is like in Kaluza-Klein theory, writing a solution of the full theory in terms of the reduced metric and vector potential 
\begin{align}
\dd s^2 &= (g_{\mu\nu} - B_{\mu\rho}g^{\rho\sigma}B_{\sigma\nu})\dd x^\mu \dd x^\nu
	+ 2B_{\mu\rho}g^{\rho\nu}\dd x^\mu \dd\tx_\nu + g^{\mu\nu}\dd\tx_\mu\dd\tx_\nu \, .
\label{eq:KKansatzDFT}  
\end{align}
By Comparing \eqref{eq:KKansatzDFT} with \eqref{eq:DFTmonopole} the reduced fields can be computed. The spacetime metric $g_{\mu\nu}$ and the non-vanishing components of the B-field $B_{\mu\nu}$ are given by
\begin{equation}
\begin{aligned}
\dd s^2 &= -\dd t^2 + \dd\vec{x}_{(5)}^{\; 2} + H(\dd z^2 + \dd\vec{y}_{(3)}^{\; 2}) \\
B_{iz} &= A_i \, .
\label{eq:NS5}
\end{aligned}
\end{equation} 
The determinant of this metric is $-H^4$ and therefore the string theory dilaton becomes
\begin{equation}
e^{-2\phi} = g^{-1/2} e^{-2d} = H^{-2}He^{-2\phi_0} = H^{-1}e^{-2\phi_0} \, .
\end{equation}
This solution is the NS5-brane solution of string theory \cite{Ortin04}, more precisely it is the NS5-brane smeared along the $z$ direction. Usually the harmonic function of the NS5-brane depends on all four transverse directions, that is $y^i$ and $z$. By smearing it over the $z$ direction the brane is no longer localized in $z$ and so the $z$-dependence is removed from the harmonic function.  

Smearing the solution along $z$ has also consequences for the field strength $H_{\mu\nu\rho}$. The NS5-brane comes with an H-flux whose only non-zero components are in the transverse directions $y^i$ and $z=y^4$. The field strength is written as
\begin{equation}
H_{mnp} = 3\partial_{[m}B_{np]} 
	= {\epsilon_{mnp}}^q\partial_q \ln H(r,z) 
\label{eq:DFTfieldstrength}
\end{equation}
where  $m=i,z=1,\dots,4$. We then note that the non-trivial part of the metric is $g_{mn}=H\delta_{mn}$ so that $g=\det g_{mn}=H^4$. This then allows us to write the field strength as
\begin{equation}
\begin{aligned}
H_{mnp} &= \sqrt{g}\tilde{\epsilon}_{mnpq}g^{qs}\partial_s \ln H \\
		&= H^2\tilde{\epsilon}_{mnpq}H^{-1}\delta^{qs}H^{-1}\partial_s H
		= \tilde{\epsilon}_{mnp}{}^q\partial_q H
\end{aligned}
\end{equation}
where the epsilon tensor has been converted to the permutation symbol (a tensor density) in order to make contact with the epsilon in a lower dimension. If the solution then is smeared along $z$, $H$ no longer depends on this coordinate. Therefore $H_{ijk} = 0$ and
\begin{equation}
\begin{aligned}
H_{ijz} &= 2\partial_{[i}B_{j]z} = \tilde{\epsilon}_{ijzk}\delta^{kl} \partial_l H \\
	&= \tilde{\epsilon}_{ijk}\delta^{kl} \partial_l H 
	= {\epsilon_{ij}}^k \partial_k H = 2\partial_{[i}A_{j]} \, .
\end{aligned}
\end{equation}
Thus the only non-zero component of the B-field (up to a gauge choice) of the smeared NS5-brane is $B_{iz}=A_i$. This then shows how the flux of the smeared NS5-brane is just the same as the usual magnetic two-form flux from a magnetic monopole for the electromagnetic field.

In conclusion, the smeared NS5-brane solution \eqref{eq:NS5} can be extracted from the DFT monopole \eqref{eq:DFTmonopole} using \eqref{eq:KKansatzDFT}. If $z$ and $\tz$ are exchanged, the same procedure recovers the KK-monopole of string theory. Since the monopole and the NS5-brane are T-dual to each other in string theory and DFT makes T-duality manifest, this should not come as a surprise. 

In order to identify the NS5-brane with the KK-monopole, it needed to be smeared along the $z$ direction. Any monopole type solution is expected to need more than a single patch to describe it (and in fact the topological charge may be viewed as the obstruction to a global description). In \cite{Papadopoulos:2014mxa} the problems of constructing a full global solution containing NS-NS magnetic flux, with patching between different local descriptions in DFT, are discussed in detail. So have we resolved those issues here?

Not really, in the case described above, because of the additional isometry in the transverse directions, the three-form flux is completely encoded in a two-form flux. (This is non-trivial and can be constructed in the usual way, \`a la Dirac). In other words because of the additional isometry $H_3=F_2\wedge dz$, so that although the $H_3$ flux is an element of the third cohomology it is really completely given by the second cohomology of which $F_2$ is a non-trivial representative.

One can now ask the question if it is possible to localize the monopole and remove this additional smearing. We will look at this next.

\subsection{The Localized Monopole Solution}
One can construct a solution which is not smeared but localized in the $z$ direction. Then the harmonic function $H$ has an explicit dependence on $z$
\begin{equation}
H(r,z) = 1 + \frac{h}{r^2 + z^2}
\end{equation} 
and the field strength $H_{\mu\nu\rho}$ in \eqref{eq:DFTfieldstrength} of the NS5-brane has two non-zero components
\begin{equation}
\begin{aligned}
H_{ijz} &= 2\partial_{[i}B_{j]z} = {\epsilon_{ij}}^k\partial_k H(r,z) = 2\partial_{[i}A_{j]} \\
H_{ijk} &= 3\partial_{[i}B_{jk]} = {\epsilon_{ijk}}\partial_z H(r,z) \, .
\end{aligned}
\end{equation}
The first one can be expressed in terms of the magnetic potential $A_i$ as before in the smeared case. The second one is new, as the $\partial_z$ derivative now does not vanish. The localized monopole solution of DFT then reads
\begin{equation}
\begin{aligned}
\dd s^2 	&= H(1+H^{-2}A^2) \dd z^2 + H^{-1} \dd \tz^2 \\
	&\quad + 2H^{-1} A_i\dd y^i \dd \tz - 2H^{-1} A^i \dd \ty_i \dd z
			+ 2H^{-1}{B_i}^j\dd y^i\dd \ty_j \\
	&\quad + H(\delta_{ij}+H^{-2}A_iA_j+H^{-2}{B_i}^kB_{kj}) \dd y^i \dd y^j 
			+ H^{-1} \delta^{ij} \dd \ty_i \dd \ty_j \\
	&\quad +\eta_{ab}\dd x^a \dd x^b + \eta^{ab}\dd \tx_a \dd \tx_b
\end{aligned}
\label{eq:DFTlocalized}
\end{equation}
where extra terms for $\dd y^2$ and $\dd y^i\dd \ty_j$ involving $B_{ij}$ arise as compared to \eqref{eq:DFTmonopole}. 

Upon rewriting this solution by using the ansatz \eqref{eq:KKansatzDFT}, one obtains the localized NS5-brane with its full field strength. If we carry out the simple operation of swapping the roles of $z$ and  $\tz$ in the reduction, then this gives the following result
\begin{equation}
\begin{aligned}
\dd s^2 &= -\dd t^2 + \dd\vec{x}_{(5)}^{\, 2} 
				+ H^{-1}\left[\dd \tz + A_i\dd y^i\right]^2 + H\dd\vec{y}_{(3)}^{\, 2} \\
H_{ijk} &= 3\partial_{[i}B_{jk]} = \epsilon_{ijk}\partial_z H(r,z) \, .
\end{aligned}
\end{equation}
This solution is the KK-monopole. The spacetime coordinates in this duality frame now include $\tz$, crucially though the harmonic function $H$ still depends on $z$, which is a dual coordinate in this frame. One thus concludes that this is the monopole localized in the dual winding space. This property is discussed in detail in \cite{Jensen:2011jna}. This is exactly the same result as blindly applying the Buscher rules (which would require an isometry) to the localized NS5-brane along the $z$ direction. It produces the monopole (which is indeed the T-dual of the fivebrane) but the solution is localized in the dual winding direction.

The alert reader will be aware that obviously one should not be allowed to use the Buscher rules to carry out a T-duality in the $z$ direction in the case where the NS5-brane is localized. The $z$ direction is not an isometry of the localized solution. Here we have a very clear example of how Double Field Theory differs from just a theory with manifest T-duality. {\bf{Double Field Theory makes no assumptions about the existence of isometries}.} The $O(d,d)$ symmetry in DFT is a local continuous symmetry that is applicable for {\it{any}} background. This perspective was discussed in \cite{Berman:2014jba} amongst other places, most recently in \cite{Cederwall:2014opa}. 

The usual spacetime manifold is defined by picking out a maximally isotropic subspace of the doubled space. Normally this is done by solving the section condition or strong constraint, which removes the dependence of fields on half of the coordinates. We then identify the remaining coordinates with the coordinates of spacetime. 

The DFT monopole is a single DFT solution which obeys the section condition; how we identify spacetime is essentially a choice of the {\it{duality frame}}. When the half-dimensional subspace which we call spacetime matches that of the reduction through the section condition, then we have a normal supergravity solution which, in the case described above, is the NS5-brane. Alternatively, one can pick the identification of spacetime not to be determined by the section condition, this then gives an alternative duality frame. Generically this will not have a supergravity description even though it is part of a good DFT solution. This is precisely the case described in this section. There is a localization in winding space and so this solution cannot be described through supergravity alone -- even though it maybe a good string background. In DFT it is just described by picking a spacetime submanifold that is not determined by the solution of the section condition.

\begin{table}[h]
\begin{center}
\begin{tabular}{|c|c|c|}
\hline
duality frame & \begin{tabular}[c]{@{}c@{}}DFT solution \\ with $H=H(r,z)$\end{tabular}          & \begin{tabular}[c]{@{}c@{}}DFT solution \\ with $H=H(r,\tz)$\end{tabular}       \\ \hline
A             & \begin{tabular}[c]{@{}c@{}}NS5-brane\\ localized in spacetime\end{tabular}       & \begin{tabular}[c]{@{}c@{}}NS5-brane \\ localized in winding space\end{tabular} \\ \hline
B             & \begin{tabular}[c]{@{}c@{}}KK-monopole\\ localized in winding space\end{tabular} & \begin{tabular}[c]{@{}c@{}}KK-monopole\\ localized in spacetime\end{tabular}    \\ \hline
\end{tabular}
\end{center}
\caption{In this table both DFT solutions are of the form \eqref{eq:DFTlocalized} but with different coordinate dependencies in the harmonic function. Each solution can be viewed in two different duality frames. In frame A the $z$ coordinate is a spacetime coordinate while $\tz$ is a dual winding coordinate. In frame B it is the other way round, $z$ is a dual winding coordinate while $\tz$ is a spacetime coordinate. The solutions extracted from the DFT solutions that are localized in spacetime have good supergravity descriptions while those that are localized in winding space have not.}
\label{tab:localized}
\end{table}

With this in mind, we come to the following conclusion. There are two different DFT solutions of the form \eqref{eq:DFTlocalized}, one with $H(r,z)$ and the other with $H(r,\tz)$ as harmonic function. Here by $z$ and $\tz$ we do not mean spacetime and winding coordinates \emph{a priori}, but just the coordinates as expressed in \eqref{eq:DFTlocalized}. For each of these two DFT solutions there is a choice of duality frames which are of course related by $O(d,d)$ rotations. In one frame, for clarity call it frame A, $z$ is a spacetime coordinate and $\tz$ is a dual winding coordinate. In another frame, say frame B, the role of $z$ and $\tz$ is exchanged, i.e. $\tz$ is a spacetime coordinate and $z$ is dual. See Table \ref{tab:localized} for an overview. 

In the case where $H$ is a function of $z$, the DFT solution rewritten in the duality frame A is the NS5-brane localized in spacetime. Its T-dual, found by going to frame B, is the KK-monopole localized in winding space which has no supergravity description as explained above. In the other case where $H$ is a function of $\tz$, the DFT solution rewritten in frame B gives the KK-monopole localized in spacetime while frame A gives the NS5-brane localized in winding space. Again this is a solution with no supergravity description but valid from a string theory point of view. 

The DFT solution listed in the first column of Table \ref{tab:localized} containing the winding localized monopole and spacetime localized NS5-brane was first given in the work by Jensen \cite{Jensen:2011jna}. The DFT solution described in the second column extend Jensen's ideas but are of course a natural consequence of the structure of DFT. We would also like to emphasize that one may interpret Jensen's solution as a DFT monopole as described here (this interpretation has not been made before).  

{\bf{What then is T-duality?}} When there is a spacetime isometry then there is indeed an ambiguity in how one identifies the spacetime in doubled space. The presence of the isometry means there are no unwanted dependences on dual coordinates from picking different duality frames and so supergravity is a good description for both choices. Thus from the DFT perspective, traditional T-duality comes from an ambiguity in how one defines the half-dimensional subspace corresponding to a good supergravity solution. This perspective of T-duality and the identification of spacetime as a null subspace, determined by the $O(d,d)$ structure $\eta$ was described first in \cite{Hull:2004in}. We do not differ from this perspective. Where we perhaps extend the description in \cite{Hull:2004in} is that DFT does allow us to pick subspaces that do not match the section condition. This choice does not allow a spacetime interpretation but does have an interpretation from string theory.

In \cite{Tong:2002rq} and more recently in related works by Harvey and Jensen \cite{Harvey:2005ab, Jensen:2011jna} and Kimura \cite{Kimura:2013fda,Kimura:2013zva,Kimura:2013khz,Kimura:2014upa,Kimura:2014bxa,Kimura:2014aja} a gauged linear sigma model was used to describe the NS5-brane and related solutions. By ``related solutions'' we mean the KK-monopole and in fact also the exotic $5_2^2$ brane \cite{deBoer:2010ud,deBoer:2012ma}. These are all solutions in the same $O(d,d)$ duality orbit. The advantage of the gauged linear sigma model description is that one may examine the inclusion of world sheet instanton effects. As first shown in \cite{Tong:2002rq}, the inclusion of such world sheet instantons gives rise exactly to the localization in dual winding space we are describing above. Thus in some sense DFT knows about world sheet instantons.

In terms of the topological questions raised by \cite{Papadopoulos:2014mxa}, the localized solution (which does not have the additional isometry) requires an appropriate patching to form a globally defined solution. Thus for this paper we restrict ourselves to giving only descriptions in a local patch. What is hopeful is that the solution described here has very specific topology of the dual space since it is itself a monopole. It is hoped to carry out a detailed analysis of the global properties in the future.

\section{The Exceptional Case $E_7$}
There are similar constructions to DFT for the U-duality groups of M-theory. In this paper we will work with the $E_7$ group. For more on this, see \cite{Berman:2011jh}. The approach described in \cite{Berman:2011jh} is in fact a truncated version of the full theory. Recently, through an excellent series of works, the full non-truncated theory, which goes by the name Exceptional Field Theory, has been developed by Hohm and Samtleben \cite{Hohm:2013pua,Hohm:2013vpa,Hohm:2013uia,Hohm:2014fxa,Godazgar:2014nqa}. We will not deal with this non-truncated version of the theory in this paper but we hope to investigate properties of solutions to the Hohm and Samtleben theory in the future \cite{Berman:2014hna}.

\subsection{The $E_7$ Exceptional Extended Geometry}
We consider the case where the eleven-dimensional theory is a direct product of $M^4 \times M^7$, the U-duality group acting on the seven-dimensional space $M^7$ is $E_7$. We will truncate the theory to ignore all dependence on the $M^4$ directions and will not allow any excitations of fields with {\it{mixed indices}} such as the graviphoton. The exceptional extended geometry is constructed by combining the seven spacetime dimensions with {\it{wrapping directions}} of the M2-brane, M5-brane and KK-monopole to form a 56-dimensional extended space with tangent space given by
\begin{equation}
TM \oplus \Lambda^2 T^*M \oplus \Lambda^5 T^*M \oplus (T^*M \otimes \Lambda^7 T^*M) \, .
\end{equation}
Details of this construction and the resulting theory are described in \cite{Hillmann:2009ci} and \cite{Hull:2007zu,Pacheco:2008ps,Coimbra:2011ky,Coimbra:2012af,Berman:2010is,Berman:2011jh,Berman:2012vc}. The algebra is $E_7\otimes GL(4)$ with the $E_7$ acting along the seven spacetime dimensions of the extended space. The generators of the associated motion group are
\begin{equation}
P_\mu, Q^{\mu\nu},Q^{\mu_1\dots\mu_5},Q^{\mu_1\dots\mu_7,\nu}\qandq P^\alpha
\end{equation}
where $\mu=1,\dots,7$ and $\alpha=1,\dots,4$. The first four generate the $\mathbf{56}$ representation of $E_7$ and the last one generates translations in the remaining four directions, the $GL(4)$. For convenience, the following dualization of generators is used
\begin{equation}
\tQ_{\mu\nu}=\frac{1}{5!}\epsilon_{\mu\nu\rho_1\dots\rho_5}Q^{\rho_1\dots\rho_5} 
\qandq
\tQ^{\mu}=\frac{1}{7!}\epsilon_{\nu_1\dots\nu_7}Q^{\nu_1\dots\nu_7,\mu} \, .
\end{equation}
For the $E_7$ generators we can now introduce generalized coordinates 
\begin{equation}
\XX^M = (X^\mu,Y_{\mu\nu},Z^{\mu\nu},W_\mu)
\end{equation}
to form the extended 56-dimensional space. Note that an index pair $\mu\nu$ is antisymmetric and we thus have indeed $7+21+21+7=56$ coordinates.

The generalized metric $\MM_{MN}$ of this extended space can be constructed from the vielbein given in \cite{Hillmann:2009ci,Hull:2007zu,Pacheco:2008ps,Coimbra:2011ky,Coimbra:2012af,Berman:2011jh}. The full expression is quite an unwieldy structure, so we will introduce it in several steps.

The underlying structure of $\MM_{MN}$ can be seen clearly if the M-theory potentials $C_3$ and $C_6$ are turned off. Then the only field present is the spacetime metric $g_{\mu\nu}$ and the line element of the extended space\footnote{As for the generalized metric in DFT, we utilize a line element to present the components of the matrix $\MM_{MN}$ and the coordinates of the extended space in a concise form. We do not wish to imply that the generalized metric is an actual metric tensor on the extended space.} reads
\begin{equation}
\begin{aligned}
\dd s^2 &= \MM_{MN}\dd\XX^M\dd\XX^N \\
		&= g^{-1/2}\Big\{
			g_{\mu\nu}\dd X^\mu X^\nu
			+ g^{\rho\sigma,\lambda\tau}\dd Y_{\rho\sigma}\dd Y_{\lambda\tau} \\
		&\qquad\qquad 
			+ g^{-1} g_{\rho\sigma,\lambda\tau}\dd Z^{\rho\sigma}\dd Z^{\lambda\tau}  
			+ g^{-1}g^{\mu\nu}\dd W_\mu \dd W_\nu \Big\} \, .
\end{aligned}
\label{eq:genmetric}
\end{equation}
Here the determinant of the spacetime metric is denoted by $g=\det g_{\mu\nu}$ and the four-index objects are defined by $g_{\mu\nu,\rho\sigma}=\frac{1}{2} \left(g_{\mu\rho}g_{\nu\sigma} - g_{\mu\sigma}g_{\nu\rho}\right)$ and similarly for the inverse. 

The generalized metric has a scaling symmetry and can be rescaled by a power of its determinant which in turn is just a power of $g$. The bare metric, i.e. without the factor of $g^{-1/2}$ upfront, has $\det \MM_{MN} = g^{-28}$. One could choose to rescale by including a factor of $g^{1/2}$ which would then lead to $\det\MM_{MN}=1$, an often useful and desirable property. 

Here the factor $g^{-1/2}$ is included. It arises completely naturally from the $E_{11}$ programme, see \cite{Berman:2011jh}, and interestingly gives solutions in the Einstein frame when rewritten by a KK-ansatz (i.e. no further rescaling is necessary).

If the gauge potentials are non-zero, there are additional terms for the ``diagonal'' entries of \eqref{eq:genmetric} and also ``cross-terms'' mixing the different types of coordinates. For what follows we will not need to use the full generalized metric with both potentials present at the same time. We will just need to consider the two special cases where either the $C_3$ potential or the $C_6$ potential vanishes. 

In the first case with no three-form, the six-form is dualized and encoded as
\begin{equation}
U^\mu = \frac{1}{6!}\epsilon^{\mu\nu_1\dots\nu_6}C_{\nu_1\dots\nu_6}
\label{eq:defU}
\end{equation}
which allows the line element to be written as
\begin{equation}
\begin{aligned}
\dd s^2 &= g^{-1/2}\left\{ \left[g_{\mu\nu} 
	+ \frac{1}{2}(g_{\mu\nu}U^\rho U_\rho - U_\mu U_\nu)\right]\dd X^\mu X^\nu \right.
	+ \frac{2}{\sqrt{2}}g^{-1/2}g_{\mu[\lambda}U_{\tau]}\dd X^\mu \dd Z^{\lambda\tau} \\
	&\qquad 
	+ \left[g^{\rho\sigma,\lambda\tau} 
		- \frac{1}{2}U^{[\rho}g^{\sigma][\lambda}U^{\tau]}\right] 
			\dd Y_{\rho\sigma}\dd Y_{\lambda\tau} 
	+ \frac{2}{\sqrt{2}}g^{-1/2}U^{[\rho}g^{\sigma]\nu} \dd Y_{\rho\sigma}\dd W_\nu \\
	&\qquad
	+ g^{-1} g_{\rho\sigma,\lambda\tau}\dd Z^{\rho\sigma}\dd Z^{\lambda\tau}  
		+ g^{-1}g^{\mu\nu}\dd W_\mu \dd W_\nu \vphantom{\frac{1}{2}} 
		\left. \vphantom{\frac{1}{2}} \right\} \, .
\end{aligned}
\label{eq:genmetric1}
\end{equation}

In the second case with no six-form, the three-form components are encoded in $C, V$ and $X$ (see \cite{Berman:2011jh}). We will concentrate on the special case where 
\begin{equation}
V^{\mu_1\dots\mu_4} = 
	\frac{1}{3!}\epsilon^{\mu_1\dots\mu_4\nu_1\dots\nu_3}C_{\nu_1\dots\nu_3}\neq 0
\quad\mathrm{but}\quad 
{X_\mu}^{\rho\sigma} = C_{\mu\lambda\tau}V^{\lambda\tau\rho\sigma} = 0 \, .
\label{eq:defV}
\end{equation}
Then the line element for the generalized metric is then given by
\begin{equation}
\begin{aligned}
\dd s^2 &= g^{-1/2}\left\{\left[g_{\mu\nu} 
		+ \frac{1}{2}C_{\mu\nu\rho}g^{\rho\sigma,\lambda\tau}
			C_{\lambda\tau\nu} \right]\dd X^\mu \dd X^\nu \right. \\ 
	&\qquad 	
		+ \left[g^{\mu_1\mu_2,\nu_1\nu_2} + \frac{1}{2}V^{\mu_1\mu_2\rho\sigma}
			g_{\rho\sigma,\lambda\tau}V^{\lambda\tau\nu_1\nu_2}\right] 
				\dd Y_{\mu_1\mu_2}\dd Y_{\nu_1\nu_2} \\
	&\qquad  + g^{-1} \left[g_{\mu_1\mu_2,\nu_1\nu_2} 
		+ \frac{1}{2}C_{\mu_1\mu_2\rho}g^{\rho\sigma}C_{\sigma\nu_1\nu_2}\right]
				\dd Z^{\mu_1\mu_2}\dd Z^{\nu_1\nu_2} \\ 
	&\qquad	
		+ g^{-1}g^{\mu\nu}\dd W_\mu \dd W_\nu 
		+ \frac{2}{\sqrt{2}}g^{-1/2}C_{\mu\rho\sigma}g^{\rho\sigma,\lambda\tau}
			\dd X^\mu \dd Y_{\lambda\tau} \\ 
	&\qquad	
		+ \frac{2}{\sqrt{2}}g^{-1/2}V^{\mu_1\mu_2\rho\sigma}g_{\rho\sigma,\nu_1\nu_2} 
			\dd Y_{\mu_1\mu_2}\dd Z^{\nu_1\nu_2}  \\ 
	&\qquad	
		+ \left.\frac{2}{\sqrt{2}}g^{-1/2}C_{\mu_1\mu_2\rho}g^{\rho\nu}
			\dd Z^{\mu_1\mu_2}\dd W_\nu  \right\} \, .
\end{aligned}
\label{eq:genmetric2}
\end{equation}

The action for the $E_7$ theory can be constructed as in \cite{Hillmann:2009ci,Hull:2007zu,Pacheco:2008ps,Coimbra:2011ky,Coimbra:2012af,Berman:2011jh}. One should remember though that when deriving the equations of motion through the variation of the action, it is necessary that the generalized metric remains in the $E_7/SU(8)$ coset. Thus the variation is subject to a constraint. This has the effect of introducing a projector on the naive equations of motion. This set of projected equations of motion was first worked out for DFT in \cite{Hohm:2010pp} and for the $SL(5)$ exceptional case in \cite{Berkeley:2014nza} along with the general formula for the exceptional cases. A solution of the exceptional extended geometry thus has to satisfy
\begin{equation}
{P_{MN}}^{KL}K_{KL} = 0
\end{equation}
where $P$ is the projector of the $E_7$ theory  and $K$ is the variation of the action with respect to the generalized metric $\MM$. (The indices are taken to run from 1 to 56 and appear in symmetric pairs.)

Before we go on to construct and discuss specific solutions to the $E_7$ theory, let as briefly recall some classic M-theory solutions. This allows us to present our conventions and clarify the notation.

\subsection{Classic Supergravity Solutions}
In eleven-dimensional supergravity there are four classic solutions: the wave, the membrane, the fivebrane and the monopole. They are all related by T- and S-duality and upon reduction on a circle they give rise to the spectrum of string theory solutions in ten dimensions.

Here we will briefly present these four solutions in terms of the bosonic fields $C_3,C_6$ and $g$ which in turn are given terms of an harmonic function $H$. To allow for easy comparison of the solutions, they are all expressed in the same coordinate system, even if is not the most natural for each solution. The coordinates we choose have one time direction $t$, one ``special'' direction $z$, six directions $\vec{x}_{(6)} = x^a$ and three directions $\vec{y}_{(3)}=y^i$ for a total of eleven dimensions. The reason for this notation will become apparent soon. 

The order of these coordinates is important for the extended coordinates with an antisymmetric pair of indices since for example $Y_{tz}=-Y_{zt}$. It is fixed by defining the permutation symbol $\epsilon_{tx^1x^2x^3x^4x^5x^6y^1y^2y^3z}=+1$. This order will be kept also after reductions when some of the coordinates drop out.

Let's start with the ``pure gravity'' solutions, the pp-wave and the KK-monopole. They do not come with a gauge potential and are given just in terms of the metric. The pp-wave consists of parallel rays carrying momentum in the $z$ direction with transverse plane wavefronts spanned by $x^a$ and $y^i$ in the above mentioned coordinates. The wave solution then reads
\begin{equation}
\begin{aligned}
\dd s^2 &= -H^{-1}\dd t^2 + H\left[\dd z - (H^{-1}-1)\dd t\right]^2 
				+ \dd\vec{x}_{(6)}^{\, 2} + \dd\vec{y}_{(3)}^{\, 2}   \\
		&= (H-2)\dd t^2 + 2(H-1)\dd t \dd z + H\dd z^2 + \delta_{ab}\dd x^a\dd x^b + \delta_{ij}\dd y^i\dd y^j \\
H &= 1 + \frac{h}{|\vec{x}_{(6)}^{\, 2} + \vec{y}_{(3)}^{\, 2} |^{7/2}}  
\end{aligned}
\label{eq:classicwave}
\end{equation}
where $h$ is some constant proportional to the momentum carried.

The KK-monopole or KK6-brane solution was already introduced in Section \ref{sec:DFT}. Where-as the momentum of the wave solution can be seen as \emph{gravito-static} charge, the monopole carries topological or \emph{gravito-magnetic} charge, hence the name ``monopole''. This solution is expressed in terms of a vector potential $A_i$ which is related to the harmonic function as before, see equation \eqref{eq:AH}. For the monopole, the $z$ direction needs to be compact and will be referred to as the ``KK-circle''. The $x^a$ form the world volume of the KK6-brane, leaving the $y^i$ to be transverse. For completeness, the monopole solution is restated in full
\begin{equation}
\begin{aligned}
\dd s^2 &= -\dd t^2 + \dd\vec{x}_{(6)}^{\, 2} 
				+ H^{-1}\left[\dd z + A_i\dd y^i\right]^2 + H\dd\vec{y}_{(3)}^{\, 2}  \\
		&= -\dd t^2 + \delta_{ab}\dd x^a\dd x^b + H^{-1}\dd z^2 + 2H^{-1}A_i\dd y^i\dd z
				+ H\left(\delta_{ij}+H^{-2}A_iA_j\right)\dd y^i\dd y^j \\
H &= 1 + \frac{h}{|\vec{y}_{(3)}|} \, , \qquad
\partial_{[i}A_{j]} = \frac{1}{2}{\epsilon_{ij}}^k\partial_k H \, .
\end{aligned}
\label{eq:classicmonopole}
\end{equation}
Again $h$ is a constant, here it is proportional to the magnetic charge.

Now turn to the extended solutions, the M2-brane and the M5-brane. These branes naturally couple to the $C_3$ and $C_6$ gauge potentials respectively. This can be seen as the natural \emph{electric} coupling. 

For both branes the worldvolume is spanned by $t$ and some of the $x^a$, while the remaining $x$'s, $y^i$ and $z$ are transverse to it. The harmonic function $H$ in each case is a function of the transverse directions. The membrane solution is given by
\begin{equation}
\begin{aligned}
\dd s^2 &= H^{-2/3}[-\dd t + \dd\vec{x}_{(2)}^{\, 2}] + H^{1/3}[\dd\vec{x}_{(4)}^{\, 2} + \dd\vec{y}_{(3)}^{\, 2} + \dd z^2]   \\
C_{tx^1x^2} &= -(H^{-1}-1), \qquad \tC_{izx^3x^4x^5x^6} = A_i\\
H &= 1 + \frac{h}{|\vec{x}_{(4)}^{\, 2} + \vec{y}_{(3)}^{\, 2} + z^2|^{6/2}} 
\end{aligned}
\label{eq:classicmembrane}
\end{equation}
and the fivebrane solution reads 
\begin{equation}
\begin{aligned}
\dd s^2 &= H^{-1/3}[-\dd t + \dd\vec{x}_{(5)}^{\, 2}] + H^{2/3}[\dd x_6^2 + \dd\vec{y}_{(3)}^{\, 2} + \dd z^2]   \\
\tC_{tx^1x^2x^3x^4x^5} &= -(H^{-1}-1),  \qquad C_{izx^6} = A_i  \\
H &= 1 + \frac{h}{|x_6^2 + \vec{y}_{(3)}^{\, 2} + z^2|^{3/2}}  \, .
\end{aligned}
\label{eq:classicfivebrane}
\end{equation}
In both cases both the electric and magnetic potentials are shown. The latter ones can be found by dualizing the corresponding field strengths. The field strength of the electric potential is proportional to $F\sim \partial H^{-1}\sim \partial H$ which is dualized into  $\tF\sim\epsilon\partial H\sim \partial A$ where we use \eqref{eq:AH} to relate $H$ and $A$. Therefore the vector potential $A_i$ appears in the components of the magnetic potentials.

The four solutions recapped above are all related to each other by M-theory dualities. The wave and the membrane are T-dual to each other, in the same way the wave and the fundamental string are related by T-duality in string theory. Similarly the monopole and the fivebrane are T-duals, again as for the monopole and NS5-brane in string theory (cf. Section \ref{sec:DFT}).

Furthermore, the membrane and fivebrane are related by S-duality, they are electromagnetic duals of each other. To complete the picture, there is a S-duality relation between the wave and the monopole. We will discuss this further towards the end of this paper. In Table \ref{tab:classicsol} the character of each of the eleven dimensions for each of the four solutions is illustrated. 
 
\begin{table}[h]
\begin{center}
\begin{tabular}{l|lllllllllll}
solution    & $t$ & $x^1$ & $x^2$ & $x^3$ & $x^4$ & $x^5$ & $x^6$ & $y^1$   & $y^2$   & $y^3$   & $z$     \\ \hline
pp-wave     & -   &       &       &       &       &       &       &         &         &         & -       \\
KK-monopole & -   & -     & -     & -     & -     & -     & -     & $\bullet$ & $\bullet$ & $\bullet$ & $\bullet$ \\
M2-brane    & -   & -     & -     &       &       &       &       & $\circ$   & $\circ$   & $\circ$   & $\circ$   \\
M5-brane    & -   & -     & -     & -     & -     & -     &       & $\circ$   & $\circ$   & $\circ$   & $\circ$
\end{tabular}
\end{center}
\caption{In this table a dash denotes that the solution is extended in that direction while a blank denotes a transverse direction. For the monopole, the four transverse directions (denoted by a dot) are special in the sense that the magnetic potential $A_i$ and the KK-circle $z$ encapsulate all the non-trivial features of the monopole. These four directions are of interest for the M2 and M5 because they are the directions (denoted by a circle) through which the electric or magnetic fluxes will flow.}
\label{tab:classicsol}
\end{table}

If these classic solutions are carried over from eleven-dimensional supergravity to the extended $E_7$ theory, the underlying spacetime has to be reduced from eleven to seven dimensions in order to build the 56-dimensional extended space. There are various ways of picking the seven and four out of the eleven as will be explained below. 

Note that in order to keep the notation simple we will use the following convention. If the directions $x^3,x^4$ and $x^5$ are reduced, we still use $x^a$ with $a=1,2$ for the first two $x$'s or alternatively label them as $x^1=u$ and $x^2=v$. Similarly we use $x^6=w$ where necessary.

\subsection{The M2- and M5-brane as a Wave in Exceptional Extended Geometry}
\label{sec:M2M5wave}
In \cite{Berkeley:2014nza} it was not only shown how the wave in DFT gives rise to the fundamental string but also that a null wave in the $SL(5)$ extended theory reduces to the membrane in ordinary spacetime. The same is true for the $E_7$ extended theory. A null wave propagating along a membrane wrapping direction gives rise to the M2-brane. 

Furthermore, due to the larger extended space, it is now also possible to consider a wave travelling in a fivebrane wrapping direction. Unsurprisingly, this reduces to the M5-brane in ordinary spacetime. We will demonstrate this explicitly and for completeness reproduce the membrane result. 

In DFT, the section condition is easily solved by reducing the coordinate dependence to half the doubled space. Thus each pair of solutions related by an $O(d,d)$ transformation, such as the wave and string or the monopole and fivebrane, can be presented in a straightforward fashion. In contrast in the exceptional extended geometry, the solutions to the section condition are more complex since a much larger extended space has to be dealt with. In the case of $E_7$, the section condition takes one from 56 to seven dimensions. We thus present the solutions step by step and relate them ``by hand'' rather than constructing the different solutions to the section condition explicitly.

Consider the following solution for an extended $E_7$ theory built from a seven-dimensio-nal spacetime with coordinates $X^\mu=(t,x^m,z)\rightarrow\XX^M$ with $m=1,\dots,5$, i.e. in the above mentioned coordinate system reduce on $x^3,x^4,x^5$ and $x^6$ and collect the remaining transverse directions $x^1,x^2$ and $y^i$ into $x^m$. The generalized metric is given by\footnote{The delta with four indices is defined as $\delta_{mn,kl}=\frac{1}{2}\left(\delta_{mk}\delta_{nl}-\delta_{ml}\delta_{nk}\right)$ and similarly for the inverse.}
\begin{equation}
\begin{aligned}
\dd s^2 &= (2-H)\left[-(\dd X^t)^2 + \delta^{mn}\dd Y_{mz} \dd Y_{nz} 
						+ \delta_{mn}\dd Z^{tm} \dd Z^{tn} - (\dd W_z)^2 \right] 
						- (\dd Y_{tz})^2\\
		&\qquad + H\left[(\dd X^z)^2 - \delta^{mn}\dd Y_{tm} \dd Y_{tn} 
						- \delta_{mn}\dd Z^{mz} \dd Z^{nz} + (\dd W_t)^2 \right]	
						+ (\dd Z^{tz})^2\\	
		&\qquad + 2(H-1)\left[\dd X^t \dd X^z - \delta^{mn}\dd Y_{tm} \dd Y_{nz} 
						+ \delta_{mn}\dd Z^{tm} \dd Z^{nz} - \dd W_t\dd W_z \right] \\	
		&\qquad + \delta_{mn} \dd X^m \dd X^n + \delta^{mn,kl}\dd Y_{mn} \dd Y_{kl} 
				- \delta_{mn,kl}\dd Z^{mn} \dd Z^{kl} - \delta^{mn} \dd W_m \dd W_n \, .
\end{aligned}
\label{eq:E7wave}
\end{equation}
This is a massless, uncharged null wave carrying momentum in the $X^z=z$ direction and $H=1+\frac{h}{|\vec{x}_{(5)}|^3}$ is a harmonic function of the transverse coordinates $x^m$. The solution is smeared over all other directions and thus there is no coordinate dependence on them. If the extra wrapping dimensions are reduced by using a Kaluza-Klein ansatz based on \eqref{eq:genmetric}, one recovers the pp-wave in M-theory in seven dimensions.

If the wave is rotated to travel in a different direction, the momentum it carries becomes the mass and charge of an extended object in the reduced picture. The different M-theory solutions obtained upon a KK-reduction of the extended wave solution pointing in various directions are summarized in Table \ref{tab:E7wave}.

\begin{table}[ht]
\begin{center}
\begin{tabular}{c|c}
\begin{tabular}[c]{@{}c@{}}direction of \\ propagation\end{tabular} & \begin{tabular}[c]{@{}c@{}}supergravity \\ solution\end{tabular} \\ \hline
$X\in TM$                                                                 & pp-wave                                                          \\
$Y\in \Lambda^2T^*M$                                                                 & M2-brane                                                         \\
$Z\in\Lambda^5 T^*M$                                                                 & M5-brane                                                         \\
$W\in\Lambda^6 TM$                                                                 & KK-monopole                                                     
\end{tabular}
\end{center}
\caption{The wave in exceptional extended geometry can propagate along any of the  extended directions giving the various classic solutions when seen from a supergravity perspective.}
\label{tab:E7wave}
\end{table}

The rotation that points the wave in the $Z^{tz}$ direction is achieved by the following swap of coordinate pairs in the above solution
\begin{equation}
\begin{aligned}
X^z &\longleftrightarrow Z^{tz} 	&
W_z &\longleftrightarrow Y_{tz} \\
X^{m} &\longleftrightarrow Z^{tm} &
W_{m} &\longleftrightarrow Y_{tm} \, .
\end{aligned}
\label{eq:wave2fivebrane}
\end{equation}
The rotated wave solution can now be rewritten by using a KK-ansatz based on the line element given in \eqref{eq:genmetric1} to remove the extra dimensions. This gives the M5-brane solution \eqref{eq:classicfivebrane} reduced to seven dimensions (and smeared over the reduced directions)
\begin{equation}
\begin{aligned}
\dd s^2 &= H^{1/5}\left[-\dd t + \dd\vec{x}_{(5)}^{\, 2} + H \dd z^2\right] \\
\tC_{tx^1x^2x^3x^4x^5} &= -(H^{-1}-1)  \\
H &= 1 + \frac{h}{z}  \, .
\end{aligned}
\end{equation}
The details of this calculation can be found in Appendix \ref{sec:appWave}.

It can also be shown that the wave in the $E_7$ extended theory pointing along one of the $Y$-directions gives the membrane from a reduced point of view. The key steps of this calculation are given here.

Start by splitting the transverse coordinates $x^m$ into $x^a$ and $y^i$ with $a=1,2$ and $i=1,2,3$ as before so that the extended space is given by $X^\mu=(t,x^a,y^i,z)\rightarrow\XX^M$. Then the wave can be rotated to point in the $Y_{x^1x^2}$ direction. This is achieved by the mapping
\begin{equation}
\begin{aligned}
X^z &\longleftrightarrow Y_{x^1x^2}  	&  
W_z &\longleftrightarrow Z^{x^1x^2}  \\
X^a &\longleftrightarrow  \epsilon^{ab}Y_{bz} & 
W_a &\longleftrightarrow  \epsilon_{ab}Z^{bz} \\
Y_{ij} &\longleftrightarrow \epsilon_{ijk}Z^{tk} & 
Z^{ij} &\longleftrightarrow \epsilon^{ijk}Y_{tk} 
\end{aligned}
\label{eq:wave2membrane}
\end{equation}
while leaving the remaining coordinates unaltered. The extended solution \eqref{eq:E7wave} then reads (recall that $x^1=u$ and $x^2=v$)
\begin{align}
\dd s^2 &= (2-H)\left[-(\dd X^t)^2 + \delta_{ab} \dd X^a \dd X^b 
				+ \delta^{ij}\dd Y_{iz} \dd Y_{jz} \right. \notag \\
		&\hspace{3cm} \left. 
				+ \delta_{ab}\dd Z^{ta} \dd Z^{tb} 
				+ \delta^{ij,kl}\dd Y_{ij} \dd Y_{kl} 
				- (\dd Z^{uv})^2 \right] - (\dd Y_{tz})^2 \notag \\
		&\quad + H\left[(\dd Y_{uv})^2 
				- \delta^{ab}\dd Y_{ta} \dd Y_{tb} 
				- \delta_{ij,kl}\dd Z^{ij} \dd Z^{kl} \right. \notag \\
		&\hspace{3cm} \left. 	
				- \delta^{ab} \dd W_a \dd W_b 
				- \delta_{ij}\dd Z^{iz} \dd Z^{jz}	
				+ (\dd W_t)^2 \right] + (\dd Z^{tz})^2   \label{eq:E7membrane}\\ 	
		&\quad + 2(H-1)\left[\dd X^t \dd Y_{uv} 
				- \dd X^u\dd Y_{tv} + \dd X^v\dd Y_{tu}
				- {\epsilon_{ij}}^k\dd Z^{ij} \dd Y_{kz} \right. \notag \\
		&\hspace{3cm} \left. 
				+ {\epsilon^{ij}}_k\dd Y_{ij} \dd Z^{kz} 
				+ \dd W_u \dd Z^{tv} - \dd W_v \dd Z^{tu}
				- \dd W_t\dd Z^{uv} \right]  \notag \\	
		&\quad + \delta^{ab}\dd Y_{az} \dd Y_{bz} + \delta_{ij} \dd X^i \dd X^j  
				+ (\dd X^z)^2 + \delta_{ij}\dd Z^{ti} \dd Z^{tj} 
				+ \delta^{ab}\delta^{ij} \dd Y_{ai} \dd Y_{bj}  \notag \\
		&\quad  - (\dd W_z)^2 
				- \delta^{ij}\dd Y_{ti} \dd Y_{tj}
				- \delta_{ab}\delta_{ij} \dd Z^{ai} \dd Z^{bj}  
				- \delta_{ab}\dd Z^{az} \dd Z^{bz} - \delta^{ij} \dd W_i \dd W_j \, .	 \notag		
\end{align}

The KK-ansatz to reduce this metric is based on the line element given in \eqref{eq:genmetric2}, it will be used again later in the monopole section, equation \eqref{eq:KKansatzC}. The procedure is the same as in the reduction calculation that yielded the fivebrane and gives 
\begin{equation}
\begin{aligned}
\dd s^2 &= H^{-2/5}\left[-\dd t + \dd\vec{x}_{(2)}^{\, 2} 
			+ H(\dd\vec{y}_{(3)}^{\, 2} + \dd z^2)\right]   \\
C_{tx^1x^2} &= -(H^{-1}-1)\\
H &= 1 + \frac{h}{\vec{y}_{(3)}^{\, 2} + z^2}
\end{aligned}
\end{equation}
which is the M2-brane solution reduced to seven dimensions (with the harmonic function smeared accordingly).

Hence, both the M2 and the M5 can be obtained from the same wave solution in the exceptional extended geometry and all branes in M-theory are just momentum modes of a null wave in the extended theory. The direction of the wave determines the type of brane (from the reduced perspective) or indeed gives a normal spacetime wave solution. From this point of view the duality transformations between the various solutions are just rotations in the extended space.

\subsection{The M5-brane as a Monopole in Exceptional Extended Geometry}
\label{sec:M5monopole}
In Section \ref{sec:DFT} we showed that the NS5-brane of string theory was the monopole solution of DFT. In this section we want to show something similar for the M5-brane in exceptional extended geometry. 

If the KK-circle of the monopole in the $E_7$ extended theory is not along a usual spacetime direction but instead along one of the novel $Y$-directions, then this produces a smeared fivebrane solution.

First, a slightly different extended space has to be constructed. Starting from eleven dimensions and reducing on $x^3,x^4,x^5$ and $t$ allows for a construction of the monopole solution in the extended space with coordinates $X^\mu=(x^a,w,y^i,z)\rightarrow\XX^M$ (where $w=x^6$) and potential $A_i$. The generalized metric is given by
\begin{align}
\dd s^2 &= (1+H^{-2}A^2)\left[\delta^{ab}\dd Y_{az}\dd Y_{bz} 
		+ (\dd Y_{wz})^2 + H^{-2}(\dd W_z)^2\right] \notag \\
	&\quad + (1+H^{-2}A_1^2)\left[(\dd X^1)^2 + H^{-2}\delta_{ab}\dd Z^{a1}\dd Z^{b1} 
		+ H^{-2}(\dd Z^{w1})^2\right] \notag\\
	&\qquad + (1+H^{-2}A_2^2)\left[\dots\right]+ (1+H^{-2}A_3^2)\left[\dots\right] \notag\\
	&\quad + (1+H^{-2}A_1^2+H^{-2}A_2^2)
		\left[H^{-1}(\dd Y_{3z})^2 + H^{-1}(\dd Z^{12})^2\right] \notag\\
	&\qquad + (1+H^{-2}A_1^2+H^{-2}A_3^2)	\left[\dots\right]
		+ (1+H^{-2}A_2^2+H^{-2}A_3^2)\left[\dots\right] \notag\\
	&\quad + 2H^{-2}A_1A_2\left[\dd X^1 \dd X^2 - H^{-1}\dd Y_{1z}\dd Y_{2z}
		+ H^{-1}\dd Z^{13}\dd Z^{23} \right. \label{eq:E7monopole}\\
	&\hspace{3cm} \left. + H^{-2}\delta_{ab}\dd Z^{a1}\dd  Z^{b2} 
		+ H^{-2}\dd Z^{w1}\dd Z^{w2}\right] \notag\\
	&\qquad + 2H^{-2}A_1A_3\left[\dots\right] + 2H^{-2}A_2A_3\left[\dots\right] \notag\\
	&\quad + 2H^{-1}A_1\left[H^{-1}(\dd X^1\dd X^z - \delta^{ab}\dd Y_{az}\dd Y_{b1}
		 - \dd Y_{wz}\dd Y_{w1})\right. \notag\\
	&\hspace{2.6cm}\left. + H^{-2}(\dd Y_{12}\dd Y_{2z} + \dd Y_{13}\dd Y_{3z}
		- \dd Z^{12}\dd Z^{2z} - \dd Z^{13}\dd Z^{3z}) \right. \notag\\
	&\hspace{2.6cm}\left. + H^{-3}(\delta_{ab}\dd Z^{a1}\dd Z^{bz}
		+ \dd Z^{w1}\dd Z^{wz} - \dd W_1\dd W_z)\right] \notag\\
	&\qquad + 2H^{-1}A_2\left[\dots\right] + 2H^{-1}A_3\left[\dots\right] \notag\\
	&\quad  + H^{-1}\left[\delta_{ab}\dd X^a\dd X^b + (\dd X^w)^2
		+ \delta^{ab} \dd Y_{aw}\dd Y_{bw} + \delta^{ab,cd}\dd Y_{ab}\dd Y_{cd}\right] \notag\\
	&\quad + H^{-2}\left[(\dd X^z)^2  + \delta^{ab}\delta^{ij}\dd Y_{ai}\dd Y_{bj}
		+ \delta^{ij} \dd Y_{wi}\dd Y_{wj}\right] \notag\\
	&\quad + H^{-3}\left[\delta^{ab}\dd W_a\dd W_b + (\dd W_w)^2  
		+ \delta_{ab} \dd Z^{aw}\dd Z^{bw} + \delta_{ab,cd}\dd Z^{ab}\dd Z^{cd} \right. \notag\\
	&\hspace{2cm} \left. + \delta^{ij,kl}\dd Y_{ij}\dd Y_{kl}
		+ \delta_{ij} \dd Z^{iz}\dd Z^{jz} \right] \notag\\
	&\quad + H^{-4}\left[\delta_{ab}\dd Z^{az}\dd Z^{bz} + (\dd Z^{wz})^2
		+ \delta^{ij} \dd W_i\dd W_{j}\right] \notag
\end{align}
where $A^2=A_iA^i=A_1^2+A_2^2+A_3^2$. The ellipsis denotes the same terms as in the line above, with the obvious cycling through the $i$ index. The harmonic function $H$ is a function of the three $y$'s and is given by $H=1+\frac{h}{|\vec{y}_{(3)}|}$. The relation between the harmonic function and the vector potential are as given in \eqref{eq:AH}.

This is a monopole\footnote{The solution is presented here in a coordinate patch and is therefore only valid locally. A global solution requires multiple patches with appropriate transition functions. For the purpose of this paper we do not require such a globally defined solution, we will restrict the calculations to a single coordinate patch.} with the KK-circle in the $X^z=z$ direction. The solution as before may be rotated such that this ``special'' direction is of a different kind. If the KK-circle is along $Y_{wz}$, a membrane wrapping direction, the solution reduces to a M5-brane smeared along $z$. This rotation is achieved by the following map (recall that $x^1=u$ and $x^2=v$)
\begin{equation}
\begin{aligned}
X^z &\longleftrightarrow -Y_{wz} 	&
W_z &\longleftrightarrow Z^{wz} \\
X^w &\longleftrightarrow Y_{uz} &
W_w &\longleftrightarrow Z^{uz} \\
Y_{uv} &\longleftrightarrow -Y_{vz} &
Z^{uv} &\longleftrightarrow Z^{vz} \\
Y_{ui} &\longleftrightarrow -Y_{iz} &
Z^{ui} &\longleftrightarrow Z^{iz} \\
Y_{vi} &\longleftrightarrow \frac{1}{2}\epsilon_{ijk}Z^{jk} &
Z^{vi} &\longleftrightarrow  \frac{1}{2}\epsilon^{ijk}Y_{jk} \, .
\end{aligned}
\label{eq:monopole2fivebrane}
\end{equation}
Using \eqref{eq:genmetric1} to read off the fields, the exceptional extended geometry monopole reduces to the M5-brane solution 
\begin{equation}
\begin{aligned}
\dd s^2 &= H^{-3/5}[\dd\vec{x}_{(2)}^{\, 2} 
	+ H(\dd w^2 + \dd\vec{y}_{(3)}^{\, 2} + \dd z^2)]  \notag \\
C_{izw} &= A_i  \\
H &= 1 + \frac{h}{|w^2 + \vec{y}_{(3)}^{\, 2} + z^2|^{3/2}} \, .
\end{aligned}
\end{equation}
The fivebrane is given in terms of its magnetic potential, i.e. to the dual gauge potential $C_3$ given in \eqref{eq:classicfivebrane}. The full calculation is shown explicitly in Appendix \ref{sec:appMonopole}.  

We have thus demonstrated how a monopole with its KK-circle along a membrane wrapping direction is identified with a (smeared) fivebrane. This is the analogous result to the KK-monopole/NS5-brane identification in DFT shown in Section \ref{sec:DFT}.

\subsection{The Situation for the Membrane}
In theory the same story should be true for the membrane. In the previous sections the wave was shown not only to give the membrane but also the fivebrane. From the same reasoning the monopole should not only give the fivebrane, but also the membrane. 

The problem is that this cannot be shown as simply as for the fivebrane in the $E_7$ truncated theory. To obtain the membrane from the monopole one has to consider its magnetic potential $C_6$ given in \eqref{eq:classicmembrane}. But this six-form has non-zero components with indices $C_{izx^3x^4x^5x^6}$, i.e. in directions which are truncated in order to construct the exceptional extended geometry. 

More technically, if the electric $C_3$ of the membrane is dualized in seven dimensions, this gives a two-form. This means that only some part of the above six-form lives in the seven-space that gets extended, the remainder lives in the other four directions. Thus it is not possible to describe the membrane this way and stay in the truncated space. This is simply a problem with the tools at our disposal, i.e. the truncated version of the $E_7$ exceptional field theory. By looking at all the relations we have built between the solutions in the extended space, it seems natural that a monopole with its KK-circle in a fivebrane wrapping direction gives a membrane. This problem then is demanding the full non-truncated EFT \cite{Hohm:2013uia} and we hope to report on this in future work \cite{Berman:2014hna}.

\section{Discussion and Outlook}
This paper has explored the role played by monopole-type solutions in Double Field Theory and its M-theory version, exceptional extended geometry. We have seen how the KK-monopole in both the doubled and the exceptional extended geometry can be identified with a fivebrane solution (NS5 and M5 respectively) in supergravity.
 
For the DFT monopole, we also examined the localized solutions. The key here is seeing how the $O(d,d)$ symmetry in DFT is not T-duality. {\bf{T-duality in DFT emerges only when one has sufficient isometries in the solution}}, something that is certainly in tune with our intuition. Without the additional isometries the $O(d,d)$ related solutions do not all have supergravity descriptions because they have a localization in the dual space. How can we understand the localization in the dual space? It has no supergravity description. From gauged linear sigma models this has been shown to be the result of world sheet instanton effects. Rather speculatively, this may indicate that DFT has some knowledge of world sheet instantons.
 
For the wave- and monopole-like solutions in the exceptional extended geometry, there are numerous directions of further investigations that one may consider. The most pressing is the need to study these solutions in the full non-truncated version of the theory, so called {\it{exceptional field theory}}, developed by Hohm and Samtleben. This will then allow us to see the relation between the wave- and monopole-like solutions which are obviously duals of each other. We need to do this in the full theory because the duality requires the Hodge star operation of the full eleven-dimensional spacetime. In other words, the truncated $E_7$ theory uses both $C_3$ and $C_6$ and treats them as independent. We know from eleven-dimensional supergravity though that there is a duality relation between these potentials, i.e. $F_4=\star F_7$. This is a crucial aspect of the story and is part of exceptional field theory, but is not seen in the truncated $E_7$ theory.

A further direction building on this work is to examine how black branes fit into the picture in exceptional extended geometries. In particular it would be good to know how the presence of the additional dimensions of the extended solutions affect the singularity structure and the origin of the black brane. This is reserved for future work.

We have seen how a single extended geometry solution may give rise to the membrane and fivebrane of M-theory. The orientation of the extended geometry solution determines the M-theory brane type. One may ask what happens if the orientation of the solution is directed along a linear combination of exceptional directions. It is clear that this may be used to describe M-theory brane bound states or equivalently branes with non-trivial background potentials. These solutions have been explored in detail in \cite{Berman:2007tf} where the solutions were constructed through a U-duality technique.

The NS5-brane in Type IIA has an interesting two-dimensional CFT description \cite{Aharony:1998ub} in the near horizon. It would be interesting to examine this DFT description of the fivebrane from some two-dimensional CFT point of view (note that the shift in the dilaton in the DFT description allows for different regions of validity as compared to the usual description).  

Finally, in \cite{Berkeley:2014nza} the dynamics of the Goldstone modes of the DFT wave solution were calculated to give the Tseytlin string. A similar Goldstone mode analysis for these exceptional extended geometry solutions would produce a U-duality covariant worldvolume description for the membrane/fivebrane. The analysis cannot work for the membrane or the fivebrane alone since they transform into each other under U-duality. It would be interesting to see exactly what are the Goldstone modes and describe their dynamics in order to describe how the extended geometry solutions relate to normal M-theory brane actions.

\section*{Acknowledgement}
DSB is partially supported  by the STFC consolidated grant ST/J000469/1 ``String Theory, Gauge Theory and Duality'' and FJR is supported by an STFC studentship. We wish to thank Paul Townsend for questions that inspired this paper and Martin Cederwall, Jeong-Hyuck Park, Malcolm Perry, Henning Samtleben and David Tong for discussions on various related topics.

\appendix
\section{Reduction of the DFT Monopole}
\label{sec:appDFT}
In this appendix it is demonstrated that the monopole solution of DFT presented in \eqref{eq:DFTmonopole} satisfies the equations of motion which can be derived from the action
\begin{equation}
S = \int \dd^{D}X e^{-2d} R
\label{eq:DFTaction}
\end{equation}
where the scalar $R$ is given by
\begin{equation}
\begin{aligned}
R 	&= \frac{1}{8}\HH^{MN}\partial_M\HH^{KL}\partial_N\HH_{KL} 
		- \frac{1}{2}\HH^{MN}\partial_M\HH^{KL}\partial_K\HH_{NL} \\
	&\quad+ 4\HH^{MN}\partial_M\partial_N d - \partial_M\partial_N\HH^{MN}
		-4\HH^{MN}\partial_M d \partial_N d + 4\partial_M\HH^{MN}\partial_N d \\
	&\quad+ \frac{1}{2}\eta^{MN}\eta^{KL}\partial_M{\EE^A}_K\partial_N{\EE^B}_L\HH_{AB} \, . 
\end{aligned}
\end{equation}
For a detailed presentation of this action and the meaning of the last line in $R$, see \cite{Berkeley:2014nza}. The full equations of motion are given in terms of a projector to take the fact into account that the generalized metric is constrained to parametrize a coset structure. The equations for $\HH_{MN}$ and $d$ are
\begin{align}
{P_{MN}}^{KL}K_{KL} &= 
	\frac{1}{2}\left(K_{MN} - \eta_{MP}\HH^{PK}K_{KL}\HH^{LQ}\eta_{QN}\right) = 0 \\
R &= 0
\end{align}
where $K_{MN}$ is the variation of the action with respect to the generalized metric
\begin{equation}
\begin{aligned}
K_{MN} &=  \frac{1}{8}\partial_M\HH^{KL}\partial_N\HH_{KL} + 2\partial_M\partial_N d \\
	&\qquad +(\partial_L-2\partial_L d) 		
		\left[\HH^{KL}\left(\partial_{(M}\HH_{N)K}
		- \frac{1}{4}\partial_K\HH_{MN}\right)\right]\\
	&\qquad  + \left(\frac{1}{4}\HH^{KL}\HH^{PQ}-\frac{1}{2}\HH^{KQ}\HH^{LP}\right)
		\partial_K\HH_{MP}\partial_L\HH_{NQ} \\
	&\qquad - \eta^{KL}\eta^{PQ}\left(\partial_K d\partial_L{\EE^A}_P 
		- \frac{1}{2}\partial_K\partial_L{\EE^A}_P\right)\HH_{(N|R}{\EE^R}_A\HH_{|M)Q} \, . 
\end{aligned}
\end{equation}
Here $\eta$ is the invariant $O(d,d)$ metric of DFT\footnote{The different meanings of the symbol $\eta$ should be clear from its indices.}. Thus one has to compute $R$ and $K_{MN}$ for the solution and show that they satisfy these equations of motion. 

Let us recall the components of the metric for our solution \eqref{eq:DFTmonopole}. In order to not confuse inverse and dual components, we will use a bar to denote a winding index and raised indices for inverse parts. We thus have the metric and its inverse
\begin{equation}
\begin{aligned}
\HH_{zz} &= H^{-1}						&	\HH^{zz} &= H(1+H^{-2}A^2)			\\
\HH_{\bz\bz} &= H(1+H^{-2}A^2)			&	\HH^{\bz\bz} &= H^{-1}				\\
\HH_{ij} &= H(\delta_{ij}+H^{-2}A_iA_j)	&	\HH^{ij} &= H^{-1}\delta^{ij}		\\
\HH_{\bi\bj} &= 	H^{-1}\delta_{\bi\bj}	&	
							\HH^{\bi\bj} &=H(\delta^{\bi\bj}+H^{-2}A^\bi A^\bj) 		\\
\HH_{zi} &= H^{-1}A_i					&	\HH^{zi} &= -H^{-1}A^i				\\
\HH_{\bz\bi} &= -H^{-1}A_\bi				&	\HH^{\bz\bi} &= H^{-1}A^\bi			\\
\HH_{ab} &= \eta_{ab} 					&	\HH^{ab} &= \eta^{ab} 				\\
\HH_{\ba\bb} &= \eta_{\ba\bb}			&	\HH^{\ba\bb} &= \eta^{\ba\bb} 
\end{aligned}
\end{equation}
and the DFT dilaton is simply
\begin{equation}
d = \phi_0 - \frac{1}{2}\ln H \, .
\end{equation}
The harmonic function $H$ is a function of $y^i$ only, independent of $z$ and any dual coordinate. Therefore the only relevant derivatives will be $\partial_i$. Furthermore, $H$ obeys the section condition and the Laplace equation. The vector $A_i$ (whose index can be freely raised by $\delta^{ij}$) is a function of $H$ and obeys the same constraints. In addition its divergence vanishes. The relation between $H$ and $A$ given in \eqref{eq:AH} will be used frequently.

Since $\HH$ and $d$ obey the section condition, the last line in both $R$ and $K_{MN}$ can be dropped as it vanishes under section. With these simplifications in mind, we can proceed to check the equations of motion. 

Start with $R$. Inserting the components of $\HH$, the first line reduces to 
\begin{equation}
\frac{1}{8}\HH^{MN}\partial_M\HH^{KL}\partial_N\HH_{KL} 
		- \frac{1}{2}\HH^{MN}\partial_M\HH^{KL}\partial_K\HH_{NL} =
-H^{-3}\delta^{mn}\partial_mH\partial_nH
\end{equation}
while the second line gives 
\begin{equation}
4\HH^{MN}\partial_M\partial_N d - \partial_M\partial_N\HH^{MN}
		-4\HH^{MN}\partial_M d \partial_N d + 4\partial_M\HH^{MN}\partial_N d = H^{-3}\delta^{mn}\partial_mH\partial_nH
\end{equation}
and we thus have $R=0$. 

Next we compute the components of $K_{MN}$. By inspection it can be seen that $K_{aM}$ and $K_{\ba M}$ vanish for any index $M$. Also $K_{z\bz}, K_{m\bn}, K_{z\bm}$ and $K_{\bz m}$ vanish trivially. The non-zero components are
\begin{equation}
\begin{aligned}
K_{mn} &= \frac{1}{4}H^{-2}\delta^{kl}\left[\partial_kA_m\partial_lA_n 
			- \delta_{mn}\partial_kH\partial_lH\right]  
			- H^{-3}\delta^{kl}A_{(m}\partial_{n)}A_k\partial_lH \\
	&\qquad	- \frac{1}{4}H^{-4}A_mA_n\delta^{kl}\delta^{pq}\partial_kA_p\partial_lA_q \\
K_{\bm\bn} &= \frac{1}{4}H^{-4}\delta^{kl}\left[\partial_kA_\bm\partial_lA_\bn 
			- \delta_{\bm\bn}\partial_kH\partial_lH\right]	\\
K_{zz} &= - \frac{1}{4}H^{-4}\delta^{kl}\delta^{pq}\partial_kA_p\partial_lA_q	\\
K_{\bz\bz} &= - \frac{1}{4}H^{-2}\delta^{kl}\delta^{pq}\partial_kA_p\partial_lA_q 
			+ H^{-3}\delta^{kl}\delta^{pq}A_p\partial_kA_q\partial_lH  \\
	&\qquad + \frac{1}{4}H^{-4}\delta^{kl}\left[A^pA^q\partial_kA_p\partial_lA_q 
			- A^2\partial_kH\partial_lH\right] \\
K_{mz} &= -\frac{1}{2}H^{-3}\delta^{kl}\left[2\partial_mA_k-\partial_kA_m\right]\partial_lH
			- \frac{1}{4}H^{-4}\delta^{kl}\delta^{pq}A_m\partial_kA_p\partial_lA_q \\
K_{\bm\bz} &= -\frac{1}{2}H^{-3}\delta^{kl}\partial_kA_\bm\partial_lH
			+ \frac{1}{4}H^{-4}\delta^{kl}\left[A_\bm\partial_kH\partial_lH 
			- \delta^{pq}A_p\partial_kA_\bm\partial_lA_q\right] \, .
\end{aligned}
\end{equation}
Now expand the projected equations of motion component-wise. For example, the $mn$ component of the equation reads
\begin{equation}
2{P_{mn}}^{KL}K_{KL} = K_{mn} 
	- \eta_{m\bm}\left[\HH^{\bm\bk}K_{\bk\bl}\HH^{\bl\bn} 
	+ \HH^{\bm\bz}K_{\bz\bz}\HH^{\bz\bn} 
	+ 2\HH^{\bm\bk}K_{\bk\bz}\HH^{\bz\bn}\right]\eta_{\bn n} \, .
\end{equation}
Inserting the components of $K_{MN}$ computed above into this expression yields zero once all terms are summed up properly. The same holds for all the other components of the equations of motion. They are thus satisfied by our solution.

It is interesting to note the action of the projector here. Whereas the general significance of the projector in the equations of motion was pointed out in \cite{Berkeley:2014nza}, it turned out that its presence was not strictly needed to show that the DFT wave was a solution as all the components of $K_{MN}$ vanished for it independently (see Appendix A of \cite{Berkeley:2014nza}). 

In contrast here for the DFT monopole, not all components of $K_{MN}$ are zero and only once the projector acts are the equations of motion satisfied. This might be due to different properties of the wave and monopole solution, the former being conformally invariant while the latter is not.

\section{Reduction of the Exceptional Extended Wave and Monopole}
\label{sec:appReduction}
In this appendix we fill in the details of how the extended solutions of the $E_7$ duality invariant theory can be rewritten by using a Kaluza-Klein ansatz to obtain solutions in ordinary spacetime.

\subsection{From Wave to Fivebrane}
\label{sec:appWave}
In Section \ref{sec:M2M5wave} it is explained how the extended wave solution can be rotated to carry momentum along a fivebrane wrapping direction. From a ordinary spacetime point of view, this is then the M5-brane solution of supergravity. Here this calculation is presented in detail. 

After the rotation \eqref{eq:wave2fivebrane}, the wave solution \eqref{eq:E7wave} reads
\begin{equation}
\begin{aligned}
\dd s^2 &= (2-H)\left[-(\dd X^t)^2 + \delta^{mn}\dd Y_{mz} \dd Y_{nz} 
						+ \delta_{mn}\dd X^m \dd X^n - (\dd Y_{tz})^2 \right] 
						- (\dd W_z)^2 \\
		&\qquad + H\left[(\dd Z^{tz})^2 - \delta^{mn}\dd W_m \dd W_n 
						- \delta_{mn}\dd Z^{mz} \dd Z^{nz} + (\dd W_t)^2 \right]	
						+ (\dd X^z)^2 \\	
		&\qquad + 2(H-1)\left[\dd X^t \dd Z^{tz} - \delta^{mn}\dd W_m \dd Y_{nz} 
						+ \delta_{mn}\dd X^m \dd Z^{nz} - \dd W_t\dd Y_{tz} \right]  \\	
		&\qquad + \delta_{mn} \dd Z^{tm} \dd Z^{tn} + \delta^{mn,kl}\dd Y_{mn} \dd Y_{kl} 
				- \delta_{mn,kl}\dd Z^{mn} \dd Z^{kl} - \delta^{mn}\dd Y_{tm} \dd Y_{tn} \, .
\end{aligned}
\label{eq:E7fivebrane}
\end{equation}
The KK-reduction ansatz to reduce the extended dimensions is based on the line element given in \eqref{eq:genmetric1}
\begin{equation}
\begin{aligned}
\dd s^2 &= g^{-1/2}\left\{\left[g_{\mu\nu} + \frac{1}{2}e^{2\gamma_1}
		\left(g_{\mu\nu}U^\rho U_\rho - U_\mu U_\nu\right)\right]\dd X^\mu X^\nu \right.  \\
	&\qquad + \left[e^{2\alpha_1}g^{\rho\sigma,\lambda\tau} 
			- \frac{1}{2}e^{2\gamma_2}U^{[\rho}g^{\sigma][\lambda}U^{\tau]}\right] 
				\dd Y_{\rho\sigma}\dd Y_{\lambda\tau} \\
	&\qquad + e^{2\alpha_2}g^{-1} g_{\rho\sigma,\lambda\tau}
				\dd Z^{\rho\sigma}\dd Z^{\lambda\tau}  
		+ e^{2\alpha_3}g^{-1}g^{\mu\nu}\dd W_\mu \dd W_\nu \\
	&\qquad + \frac{2}{\sqrt{2}}e^{2\beta_1}g^{-1/2}
			g_{\mu[\lambda}U_{\tau]}\dd X^\mu \dd Z^{\lambda\tau}
	+ \frac{2}{\sqrt{2}}e^{2\beta_2}g^{-1/2}U^{[\rho}g^{\sigma]\nu} 
				\dd Y_{\rho\sigma}\dd W_\nu
		\left. \vphantom{\frac{1}{2}}\right\} 
\end{aligned}
\label{eq:KKansatzU}
\end{equation}
where the scale factors $e^{2\alpha}$, $e^{2\beta}$ and $e^{2\gamma}$ are undetermined. They arise naturally in such a reduction ansatz which attempts to reduce 49 dimensions at once and will be determined by consistency.

By comparing \eqref{eq:KKansatzU} to \eqref{eq:E7fivebrane} term by term, one can step by step work out the fields of the reduced solution. The term with $\dd W^2$ gives
\begin{align}
e^{2\alpha_3}g^{-3/2}g^{zz} &= -1 &
e^{2\alpha_3}g^{-3/2}g^{tt} &= H &
e^{2\alpha_3}g^{-3/2}g^{mn} &= -H\delta^{mn}
\label{eq:WreductionGinv}
\end{align}
while the $\dd Z^2$ term gives
\begin{equation}
\begin{aligned}
e^{2\alpha_2}g^{-3/2}g_{tz,tz} &= H \, , & \qquad
e^{2\alpha_2}g^{-3/2}g_{zm,zn} &= -H\delta_{mn} \\
e^{2\alpha_2}g^{-3/2}g_{tm,tn} &= \delta_{mn}\, , & \qquad
e^{2\alpha_2}g^{-3/2}g_{mn,kl} &= -\delta_{mn,kl} \, .
\end{aligned}
\label{eq:WreductionGG}
\end{equation}
Using \eqref{eq:WreductionGinv}, the cross-term $\dd Y\dd W$ gives an expression for $U^\mu$ which encodes the six-form potential
\begin{equation}
\left.
\begin{aligned}
-e^{2\beta_2}g^{-1}U^zg^{tt} &= -(H-1) \\
-e^{2\beta_2}g^{-1}U^zg^{mn} &= (H-1)\delta^{mn}
\end{aligned}
\right\}  \quad \longrightarrow \quad
e^{2\beta_2-2\alpha_3}g^{1/2}U^z = \frac{H-1}{H}  \, .
\label{eq:WreductionU}
\end{equation}

Next consider the $\dd Y^2$ term which gives
\begin{equation}
\begin{aligned}
e^{2\alpha_1}g^{-1/2}g^{mz,nz} + e^{2\gamma_2}g^{-1/2}g^{mn}U^zU^z 
	&= (2-H)\delta^{mn}\, , & \quad
e^{2\alpha_1}g^{-1/2}g^{mn,kl} &= \delta^{mn,kl}  \\ 
e^{2\alpha_1}g^{-1/2}g^{tz,tz} + e^{2\gamma_1}g^{-1/2}g^{tt}U^zU^z 
	&= -(2-H)\, , &  \quad
e^{2\alpha_1}g^{-1/2}g^{tm,tn} &= -\delta^{mn} 
\end{aligned}
\end{equation}
and using \eqref{eq:WreductionGinv} and \eqref{eq:WreductionU} one can extract
\begin{equation}
\begin{aligned}
e^{2\alpha_1}g^{-1/2}g^{zm,zn} &= 
	\left[(2-H) + H\frac{(H-1)^2}{H^2}e^{2\gamma_2+2\alpha_3-4\beta_2}\right]\delta^{mn} = H^{-1}\delta^{mn} \\
e^{2\alpha_1}g^{-1/2}g^{tz,tz} &= 
	-\left[(2-H) + H\frac{(H-1)^2}{H^2}e^{2\gamma_2+2\alpha_3-4\beta_2}\right] = -H^{-1} 
\end{aligned}
\end{equation}
if the factor $e^{2\gamma_2+2\alpha_3-4\beta_2}$ is equal to 1. The penultimate step is to look at the $\dd X\dd Z$ term 
\begin{align}
e^{2\beta_1}g^{-1}g_{tt}U_z &= (H-1) \, , & 
e^{2\beta_1}g^{-1}g_{mn}U_z &= -(H-1)\delta_{mn} 
\end{align}
and the $\dd X^2$ term which gives 
\begin{equation}
\begin{aligned}
g^{-1/2}g_{tt} + e^{2\gamma_1}g^{-1/2}g_{tt}U^zU_z &= -(2-H) \\
g^{-1/2}g_{mn} + e^{2\gamma_1}g^{-1/2}g_{mn}U^zU_z &= (2-H) \\
g^{-1/2}g_{zz} &= 1 \, .
\end{aligned}
\end{equation}
They can all be combined to determine the two remaining components of the metric 
\begin{equation}
\begin{aligned}
g^{-1/2}g_{tt} &= -\left[(2-H) + \frac{(H-1)^2}{H}e^{2\gamma_1+2\alpha_3-2\beta_1-2\beta_2}\right] 
	= -H^{-1} \\
g^{-1/2}g_{mn} &= \left[(2-H) + \frac{(H-1)^2}{H}e^{2\gamma_1+2\alpha_3-2\beta_1-2\beta_2}\right]\delta_{mn} 
	= H^{-1}\delta_{mn} 
\end{aligned}
\end{equation}
provided that $e^{2\gamma_1+2\alpha_3-2\beta_1-2\beta_2}=1$. Collecting all the above results, we have\footnote{The order of the entries in the diagonal matrices have indices $[t,m,z]$ for $g_{\mu\nu}$ and $g^{\mu\nu}$. For $g_{\mu\nu,\rho\sigma}$ and $g^{\mu\nu,\rho\sigma}$ the order is $[tm,tz,mn,mz]$.} 
\begin{equation}
\begin{aligned}
g^{-1/2}g_{\mu\nu} &= H^{-1}\diag[-1,\delta_{mn},H] \\
e^{2\alpha_3}g^{-3/2}g^{\mu\nu} &= -H\diag[-1,\delta^{mn},H^{-1}] \\
e^{2\alpha_2}g^{-3/2}g_{\mu\nu,\rho\sigma} &= -\diag[-\delta_{mn},-H,\delta_{mn,kl},H\delta_{mn}] \\
e^{2\alpha_1}g^{-1/2}g^{\mu\nu,\rho\sigma} &= \diag[-\delta^{mn},-H^{-1},\delta^{mn,kl},H^{-1}\delta^{mn}] \, .
\end{aligned}
\end{equation}
From the first line the determinant of the spacetime metric can be computed as $g=-H^{12/5}$ and thus $g_{\mu\nu}$ is finally determined. The three objects in the other lines, the inverse metric $g^{\mu\nu}$, $g_{\mu\nu,\rho\sigma}$ and $g^{\mu\nu,\rho\sigma}$, are all related to the metric. For this to be consistent and the constraints mentioned above to be satisfied, the factors $e^{2\alpha}$, $e^{2\beta}$ and $e^{2\gamma}$ have to be
\begin{equation}
\begin{aligned}
e^{2\alpha_1} &= H^{8/5} = |g|^{2/3} & \qquad
e^{2\beta_1} &= H^{2} = |g|^{5/6} & \qquad
e^{2\gamma_1} &= H^{4/5} = |g|^{1/3} \\
e^{2\alpha_2} &= H^{16/5} = |g|^{4/3} & \qquad
e^{2\beta_2} &= H^{18/5} = |g|^{3/2} & \qquad
e^{2\gamma_2} &= H^{12/5} = |g| \\
e^{2\alpha_3} &= H^{24/5} = |g|^{2}\, . & 
\end{aligned}
\end{equation} 
With this the factor in front of $U^z$ in \eqref{eq:WreductionU} now also vanishes and the six-form potential can be worked out from \eqref{eq:defU} as
\begin{equation}
U^z = \frac{H-1}{H} \quad \longrightarrow \quad
\tC_{tx^1x^2x^3x^4x^5} = \frac{H-1}{H} = -(H^{-1}-1) \, .
\end{equation}
Thus the result of reducing the full solution \eqref{eq:E7fivebrane} down to seven dimensions is
\begin{equation}
\begin{aligned}
\dd s^2 &= H^{1/5}\left[-\dd t + \dd\vec{x}_{(5)}^{\, 2} + H \dd z^2\right] \\
\tC_{tx^1x^2x^3x^4x^5} &= -(H^{-1}-1)  \\
H &= 1 + \frac{h}{z}  \, .
\end{aligned}
\end{equation}
where the harmonic function has to be smeared over the reduced directions. This is precisely the fivebrane solution in seven dimensions, obtained from reducing \eqref{eq:classicfivebrane} on $x^3,x^4,x^5$ and $x^6$ (and smearing $H$).

\subsection{From Monopole to Fivebrane}
\label{sec:appMonopole}
In Section \ref{sec:M5monopole} the extended monopole solution with its KK-circle in a membrane wrapping direction was shown to give the fivebrane coupled to its magnetic potential in ordinary spacetime. The details of this calculation are given here.

The monopole solution \eqref{eq:E7monopole} is transformed by \eqref{eq:monopole2fivebrane} to have its KK-circle along $Y_{wz}$. The extended line element then reads
\begin{equation}
\begin{aligned}
\dd s^2 &= (1+H^{-2}A^2)\left[(X^w)^2 + (\dd Y_{uv})^2
		+ (\dd X^z)^2 + H^{-2}(\dd Z^{wz})^2\right] \\
	&\quad + (1+H^{-2}A_1^2)\left[(\dd X^1)^2 + H^{-2}(\dd Z^{1z})^2 
		+ H^{-2}(\dd Y_{23})^2 + H^{-2}(\dd Z^{w1})^2\right] \\
	&\qquad + (1+H^{-2}A_2^2)\left[\dots\right]+ (1+H^{-2}A_3^2)\left[\dots\right] \\
	&\quad + (1+H^{-2}A_1^2+H^{-2}A_2^2)
		\left[H^{-1}(\dd Y_{u3})^2 + H^{-1}(\dd Y_{v3})^2\right] \\
	&\qquad + (1+H^{-2}A_1^2+H^{-2}A_3^2)	\left[\dots\right]
		+ (1+H^{-2}A_2^2+H^{-2}A_3^2)\left[\dots\right] \\
	&\quad + 2H^{-2}A_1A_2\left[\dd X^1 \dd X^2 - H^{-1}\dd Y_{u1}\dd Y_{u2}
		- H^{-1}\dd Y_{v1}\dd Y_{v2} \right. \\
	&\hspace{3cm} \left. + H^{-2}\dd Z^{1z}\dd Z^{2z} - H^{-2}\dd Y_{12}\dd Y_{23}
		+ H^{-2}\dd Z^{w1}\dd Z^{w2}\right] \\
	&\qquad + 2H^{-2}A_1A_3\left[\dots\right] + 2H^{-2}A_2A_3\left[\dots\right] \\
	&\quad + 2H^{-1}A_1\left[H^{-1}(-\dd X^1\dd Y_{wz} + \dd X^w\dd Y_{1z} 
		+ \dd Y_{uv}\dd Z^{23} + \dd X^z\dd Y_{w1})\right. \\
	&\hspace{2.6cm}\left. + H^{-2}(-\dd Z^{v3}\dd Y_{u2} + \dd Z^{v2}\dd Y_{u3}
		- \dd Y_{v3}\dd Z^{u2} + \dd Y_{v2}\dd Z^{u3}) \right. \\
	&\hspace{2.6cm}\left. + H^{-3}(\dd Z^{1z}\dd W_w + \dd Y_{23}\dd Z^{uv}
		+ \dd Z^{w1}\dd W_z - \dd W_1\dd Z^{wz})\right] \\
	&\qquad + 2H^{-1}A_2\left[\dots\right] + 2H^{-1}A_3\left[\dots\right] \\
	&\quad  + H^{-1}\left[\delta_{ab}\dd X^a\dd X^b + (\dd Y_{uz})^2
		+ \delta^{ab} \dd Y_{aw}\dd Y_{bw} + (\dd Y_{vz})^2\right] \\
	&\quad + H^{-2}\left[(\dd Y_{wz})^2  + \delta^{ij}\dd Y_{iz}\dd Y_{jz} 
		+ \delta_{ij,kl}\dd Z^{ij}\dd Z^{kl} + \delta^{ij} \dd Y_{wi}\dd Y_{wj}\right] \\
	&\quad + H^{-3}\left[\delta^{ab}\dd W_a\dd W_b + (\dd Z^{uz})^2  
		+ \delta_{ab} \dd Z^{aw}\dd Z^{bw} + (\dd Z^{vz})^2 \right. \\
	&\hspace{2cm} \left. + \delta_{ij} \dd Z^{ui}\dd Z^{uj} 
		+ \delta_{ij} \dd Z^{vi}\dd Z^{vj}\right] \\
	&\quad + H^{-4}\left[(\dd W_w)^2 + (\dd Z^{uv})^2 + (\dd W_z)^2
		+ \delta^{ij} \dd W_i\dd W_{j}\right] \, .
\end{aligned}
\label{eq:E7fivebrane2}
\end{equation}
A suitable KK-ansatz to extract the spacetime metric and three-form potential is based on \eqref{eq:genmetric2}
\begin{equation}
\begin{aligned}
\dd s^2 &= g^{-1/2}\left\{\left[g_{\mu\nu} 
		+ \frac{1}{2}e^{2\gamma_1}C_{\mu\rho\sigma}g^{\rho\sigma,\lambda\tau}
			C_{\lambda\tau\nu} \right]\dd X^\mu \dd X^\nu \right. \\ 
	&\qquad 	
		+ \left[e^{2\alpha_1}g^{\mu_1\mu_2,\nu_1\nu_2} 
		+ \frac{1}{2}e^{2\gamma_2}V^{\mu_1\mu_2\rho\sigma}
			g_{\rho\sigma,\lambda\tau}V^{\lambda\tau\nu_1\nu_2}\right] 
				\dd Y_{\mu_1\mu_2}\dd Y_{\nu_1\nu_2} \\
	&\qquad  + g^{-1} \left[e^{2\alpha_2}g_{\mu_1\mu_2,\nu_1\nu_2} 
		+ \frac{1}{2}e^{2\gamma_3}C_{\mu_1\mu_2\rho}g^{\rho\sigma}C_{\sigma\nu_1\nu_2}\right]
				\dd Z^{\mu_1\mu_2}\dd Z^{\nu_1\nu_2} \\ 
	&\qquad
		+ e^{2\alpha_3}g^{-1}g^{\mu\nu}\dd W_\mu \dd W_\nu \\ 
	&\qquad	
		+ \frac{2}{\sqrt{2}}e^{2\beta_1}C_{\mu\rho\sigma}g^{\rho\sigma,\lambda\tau}
			\dd X^\mu \dd Y_{\lambda\tau} \\ 
	&\qquad	
		+ \frac{2}{\sqrt{2}}e^{2\beta_2}g^{-1/2}
			V^{\mu_1\mu_2\rho\sigma}g_{\rho\sigma,\nu_1\nu_2} 
			\dd Y_{\mu_1\mu_2}\dd Z^{\nu_1\nu_2}  \\ 
	&\qquad \left.
		+ \frac{2}{\sqrt{2}}e^{2\beta_3}g^{-1/2}
			C_{\mu_1\mu_2\rho}g^{\rho\nu}
			\dd Z^{\mu_1\mu_2}\dd W_\nu   \right\}
\end{aligned}
\label{eq:KKansatzC}
\end{equation}
where again the a priori undetermined scale factors $e^{2\alpha}$, $e^{2\beta}$ and $e^{2\gamma}$ have to be included. We now proceed in the usual way, comparing \eqref{eq:KKansatzC} to \eqref{eq:E7fivebrane2} term by term to determine all the fields. The scale factors are then picked to ensure a consistent solution. Start with the $\dd W^2$ term
\begin{equation}
\begin{aligned}
e^{2\alpha_3}g^{-3/2}g^{ab} &= H^{-3}\delta^{ab}\, , & \quad
e^{2\alpha_3}g^{-3/2}g^{ww} &= H^{-4} \\
e^{2\alpha_3}g^{-3/2}g^{ij} &= H^{-4}\delta^{ij}\, , & \quad
e^{2\alpha_3}g^{-3/2}g^{zz} &= H^{-4}
\end{aligned}
\label{eq:MreductionGinv}
\end{equation}
which can be used in the $\dd Z\dd W$ term to find an expression for the three-form potential 
\begin{equation}
\left.
\begin{aligned}
e^{2\beta_3}g^{-1}C_{wzi}g^{ij} &= -H^{-3}A^j \\
e^{2\beta_3}g^{-1}C_{izw}g^{ww} &= H^{-3}A_i \\
e^{2\beta_3}g^{-1}C_{wiz}g^{zz} &= H^{-3}A_i
\end{aligned}
\right\} \quad  \longrightarrow \quad
e^{2\beta_3-2\alpha_3}g^{1/2}C_{izw} = A_i \, .
\label{eq:MreductionC}
\end{equation}
Once this is established, it can be used in the $\dd Z^2$ terms 
\begin{equation}
\begin{aligned}
e^{2\alpha_2}g^{-3/2}g_{wz,wz} + e^{2\gamma_3}g^{-1/2}C_{wzi}g^{ij}C_{jwz} &= H^{-2}+H^{-4}A^2 \\
e^{2\alpha_2}g^{-3/2}g_{wi,wj} + e^{2\gamma_3}g^{-1/2}C_{wiz}g^{zz}C_{zwj} 
	&= H^{-2}\delta_{ij}+H^{-4}A_iA_j \\
e^{2\alpha_2}g^{-3/2}g_{iz,jz} + e^{2\gamma_3}g^{-1/2}C_{iz}g^{ww}C_{wjz} 
	&= H^{-2}\delta_{ij}+H^{-4}A_iA_j
\end{aligned}
\end{equation}	
together with \eqref{eq:MreductionGinv} to find
\begin{equation}
\begin{aligned}
e^{2\alpha_2}g^{-3/2}g_{wz,wz} &= H^{-2} + H^{-4}A^2 
	- e^{2\gamma_3+2\alpha_3-4\beta_3}H^{-4}A^2 = H^{-2} \\
e^{2\alpha_2}g^{-3/2}g_{wi,wj} &= H^{-2}\delta_{ij}+H^{-4}A_iA_j
	- e^{2\gamma_3+2\alpha_3-4\beta_3}H^{-4}A_iA_j = H^{-2}\delta_{ij} \\
e^{2\alpha_2}g^{-3/2}g_{iz,jz} &= H^{-2}\delta_{ij}+H^{-4}A_iA_j
	- e^{2\gamma_3+2\alpha_3-4\beta_3}H^{-4}A_iA_j = H^{-2}\delta_{ij}
\end{aligned}
\end{equation}
provided that $e^{2\gamma_3+2\alpha_3-4\beta_3}$ is equal to 1. The remaining components of $g_{\mu\nu,\rho\sigma}$ are
\begin{equation}
\begin{aligned}
e^{2\alpha_2}g^{-3/2}g_{ij,kl} &= H^{-2}\delta_{ij,kl} \, , & \quad
e^{2\alpha_2}g^{-3/2}g_{ai,bj} &= H^{-3}\delta_{ab}\delta_{ij} \\
e^{2\alpha_2}g^{-3/2}g_{aw,bw} &= H^{-3}\delta_{ab} \, , & \quad
e^{2\alpha_2}g^{-3/2}g_{az,bz} &= H^{-3}\delta_{ab} \\
e^{2\alpha_2}g^{-3/2}g_{uv,uv} &= H^{-4} \, .
\end{aligned}
\label{eq:MreductionGG}
\end{equation}	
We continue with the $\dd Y\dd Z$ terms containing the object $V^{\mu\nu\rho\sigma}$. They are all of the same form (up to a sign), for example
\begin{equation}
e^{2\beta_2}g^{-1}V^{u2v3}g_{v3,v3} = -H^{-3}A_1 \quad \longrightarrow \quad
e^{2\beta_2-2\alpha_2}g^{1/2}V^{u2v3} = -A_1
\end{equation}
where \eqref{eq:MreductionGG} was used. Looking at all the terms with the relevant sign and taking the order of the $i$-type index into account, the general expression is  
\begin{equation}
e^{2\beta_2-2\alpha_2}g^{1/2}V^{uvij} = \epsilon^{ijk}A_k  \, .
\label{eq:MreductionV}
\end{equation} 
This can in turn be used in the $\dd Y^2$ terms
\begin{equation}
\begin{aligned}
e^{2\alpha_1}g^{-1/2}g^{uv,uv} + e^{2\gamma_2}g^{-1/2}V^{uvij}g_{ij,kl}V^{kluv} &= 		
	1+H^{-2}A^2 \\
e^{2\alpha_1}g^{-1/2}g^{23,23} + e^{2\gamma_2}g^{-1/2}V^{23uv}g_{uv,uv}V^{uv23} &= 
	H^{-2}  + H^{-4}A_1^2 \\
e^{2\alpha_1}g^{-1/2}g^{a3,b3} + e^{2\gamma_2}g^{-1/2}V^{a3ci}g_{ci,dj}V^{djb3} &= 
	H^{-1}\delta^{ab} + H^{-3}\delta^{ab}(A_1^2+A_2^2)
\end{aligned}
\end{equation}
together with \eqref{eq:MreductionGG} to find
\begin{equation}
\begin{aligned}
e^{2\alpha_1}g^{-1/2}g^{uv,uv} &=   		
	1+H^{-2}A^2 - e^{2\gamma_2+2\alpha_2-4\beta_2}H^{-2}A^2 = 1 \\
e^{2\alpha_1}g^{-1/2}g^{23,23} &=
  H^{-2} + H^{-4}A_1^2 - e^{2\gamma_2+2\alpha_2-4\beta_2}H^{-4}A_1^2 = H^{-2} \\
e^{2\alpha_1}g^{-1/2}g^{a3,b3} &= 
	H^{-1}\delta^{ab} + H^{-3}\delta^{ab}(A_1^2+A_2^2) - e^{2\gamma_2+2\alpha_2-4\beta_2}H^{-3}\delta^{ab} (A_1^2+A_2^2) \\
	&= H^{-1}\delta^{ab}
\end{aligned}
\end{equation}
provided that $e^{2\gamma_2+2\alpha_2-4\beta_2}$ is equal to 1. The same holds for other values of the $i$-type index. The remaining components of $g^{\mu\nu,\rho\sigma}$ are
\begin{equation}
\begin{aligned}
e^{2\alpha_1}g^{-1/2}g^{aw,bw} &= H^{-1}\delta^{ab}  \, , & \quad
e^{2\alpha_1}g^{-1/2}g^{az,bz} &= H^{-1}\delta^{ab} \\
e^{2\alpha_1}g^{-1/2}g^{wi,wj} &= H^{-2}\delta^{ij}  \, , & \quad
e^{2\alpha_1}g^{-1/2}g^{iz,jz} &= H^{-2}\delta^{ij} \\
e^{2\alpha_1}g^{-1/2}g^{wz,wz} &= H^{-2}  \, .
\end{aligned}
\label{eq:MreductionGGinv}
\end{equation}

The final cross-term to consider is the $\dd X\dd Y$ term which together with \eqref{eq:MreductionGGinv} yields another expression for the three-form potential
\begin{equation}
\left.
\begin{aligned}
e^{2\beta_1}g^{-1/2}C_{iwz}g^{wz,wz} &= -H^{-2}A_i \\
e^{2\beta_1}g^{-1/2}C_{wiz}g^{iz,jz} &= H^{-3}A^j \\
e^{2\beta_1}g^{-1/2}C_{zwi}g^{wi,wj} &= H^{-3}A^j
\end{aligned}
\right\} \quad  \longrightarrow \quad
e^{2\beta_1-2\alpha_1}g^{1/2}C_{izw} = A_i \, .
\label{eq:MreductionCC}
\end{equation}
In a last step, the $\dd X^2$ terms 
\begin{equation}
\begin{aligned}
g^{-1/2}g_{ww} + e^{2\gamma_1}g^{-1/2}C_{wiz}g^{iz,jz}C_{jzw} &= 1+H^{-2}A^2 \\
g^{-1/2}g_{zz} + e^{2\gamma_1}g^{-1/2}C_{wiz}g^{wi,wj}C_{wjz} &= 1+H^{-2}A^2 \\
g^{-1/2}g_{ij} + e^{2\gamma_1}g^{-1/2}C_{iwz}g^{wz,wz}C_{wzj} &= \delta_{ij}+H^{-2}A_iA_j \\
g^{-1/2}g_{ab} &= H^{-1}
\end{aligned}
\end{equation}
are combined with previous statements to to determine the spacetime metric
\begin{equation}
\begin{aligned}
g^{-1/2}g_{ww} &= 1+H^{-2}A^2 - e^{2\gamma_1+2\alpha_1-4\beta_1}H^{-2}A^2 = 1 \\
g^{-1/2}g_{zz} &= 1+H^{-2}A^2 - e^{2\gamma_1+2\alpha_1-4\beta_1}H^{-2}A^2 = 1 \\
g^{-1/2}g_{ij} &= \delta_{ij}+H^{-2}A_iA_j - e^{2\gamma_1+2\alpha_1-4\beta_1}A_iA_j = \delta_{ij}
\end{aligned}
\end{equation}
provided that $e^{2\gamma_1+2\alpha_1-4\beta_1}$ is equal to 1. Collecting all the above results, we have\footnote{The order of the entries in the diagonal matrices have indices $[a,w,i,z]$ for $g_{\mu\nu}$ and $g^{\mu\nu}$. For $g_{\mu\nu,\rho\sigma}$ and $g^{\mu\nu,\rho\sigma}$ the order is $[ab,aw,ai,az,wi,wz,ij,iz]$.} 
\begin{equation}
\begin{aligned}
g^{-1/2}g_{\mu\nu} &= H^{-1}\diag[\delta_{ab},H,H\delta_{ij},H] \\
e^{2\alpha_3}g^{-3/2}g^{\mu\nu} 
	&= H^{-3}\diag[\delta^{ab},H^{-1},H^{-1}\delta^{ij},H^{-1}] \\
e^{2\alpha_2}g^{-3/2}g_{\mu\nu,\rho\sigma} 
	&= H^{-4}\diag[1,H\delta_{ab},H\delta_{ab}\delta_{ij},H\delta_{ab}, 
		H^2\delta_{ij},H^2,H^2\delta_{ij,kl},H^2\delta_{ij}] \\
e^{2\alpha_1}g^{-1/2}g^{\mu\nu,\rho\sigma} 
	&= \diag[1,H^{-1}\delta^{ab},H^{-1}\delta^{ab}\delta^{ij},H^{-1}\delta^{ab}, \\
	&\hspace{5cm}	H^{-2}\delta^{ij},H^{-2},H^{-2}\delta^{ij,kl},H^{-2}\delta^{ij}] \, .
\end{aligned}
\end{equation}
From the first line the determinant of the spacetime metric can be computed as $g=H^{4/5}$ and thus $g_{\mu\nu}$ is finally determined. The three objects in the other lines, the inverse metric $g^{\mu\nu}$, $g_{\mu\nu,\rho\sigma}$ and $g^{\mu\nu,\rho\sigma}$, are all related to the metric. For this to be consistent and the constraints mentioned above to be satisfied, the factors $e^{2\alpha}$, $e^{2\beta}$ and $e^{2\gamma}$ have to be
\begin{equation}
\begin{aligned}
e^{2\alpha_1} &= H^{-4/5} = g^{-1} & \qquad
e^{2\beta_1} &= H^{-6/5} = g^{-3/2} & \qquad
e^{2\gamma_1} &= H^{-8/5} = g^{-2} \\
e^{2\alpha_2} &= H^{-8/5} = g^{-2} & \qquad
e^{2\beta_2} &= H^{-10/5} = g^{-5/2} & \qquad
e^{2\gamma_2} &= H^{-12/5} = g^{-3} \\
e^{2\alpha_3} &= H^{-12/5} = g^{-3} & \qquad
e^{2\beta_3} &= H^{-14/5} = g^{-7/2} & \qquad
e^{2\gamma_3} &= H^{-16/5} = g^{-4} \, .  
\end{aligned}
\end{equation}
Having set the scale factors, the prefactors in \eqref{eq:MreductionC}, \eqref{eq:MreductionV} and \eqref{eq:MreductionCC} vanish and $V^{\mu\nu\rho\sigma}$ can be converted into $C_{\mu\nu\rho}$ via \eqref{eq:defV} which all boils down to $C_{izw}=A_i$. Thus the result of reducing the full solution \eqref{eq:E7fivebrane2} down to seven dimensions is
\begin{equation}
\begin{aligned}
\dd s^2 &= H^{-3/5}[\dd\vec{x}_{(2)}^{\, 2} 
	+ H(\dd w^2 + \dd\vec{y}_{(3)}^{\, 2} + \dd z^2)]  \\
C_{izw} &= A_i  \\
H &= 1 + \frac{h}{|w^2 + \vec{y}_{(3)}^{\, 2} + z^2|^{3/2}} \, .
\end{aligned}
\end{equation}
where the harmonic function is smeared over the reduced directions. This is precisely the fivebrane solution in seven dimensions, obtained from reducing \eqref{eq:classicfivebrane} on $x^3,x^4,x^5$ and $t$ (and smearing $H$) with its magnetic potential.

\addcontentsline{toc}{section}{References}
\bibliographystyle{JHEP} 
\bibliography{mybib}

\providecommand{\href}[2]{#2}\begingroup\raggedright\begin{thebibliography}{10}

\bibitem{Duff90a}
M.~Duff, {\it {Duality Rotations in String Theory}},  {\em Nucl.Phys.} {\bf
  B335} (1990) 610.

\bibitem{Duff90b}
M.~Duff and J.~Lu, {\it {Duality Rotations in Membrane Theory}},  {\em
  Nucl.Phys.} {\bf B347} (1990) 394--419.

\bibitem{Tseytlin90}
A.~A. Tseytlin, {\it {Duality Symmetric Formulation of String World Sheet
  Dynamics}},  {\em Phys.Lett.} {\bf B242} (1990) 163--174.

\bibitem{Tseytlin91}
A.~A. Tseytlin, {\it {Duality Symmetric Closed String Theory and Interacting
  Chiral Scalars}},  {\em Nucl.Phys.} {\bf B350} (1991) 395--440.

\bibitem{Siegel93a}
W.~Siegel, {\it {Two Vierbein Formalism for String Inspired Axionic Gravity}},
  {\em Phys.Rev.} {\bf D47} (1993) 5453--5459,
  [\href{http://xxx.lanl.gov/abs/hep-th/9302036}{{\tt hep-th/9302036}}].

\bibitem{Siegel93b}
W.~Siegel, {\it {Superspace Duality in Low-Energy Superstrings}},  {\em
  Phys.Rev.} {\bf D48} (1993) 2826--2837,
  [\href{http://xxx.lanl.gov/abs/hep-th/9305073}{{\tt hep-th/9305073}}].

\bibitem{Siegel93c}
W.~Siegel, {\it {Manifest duality in low-energy superstrings}},
  \href{http://xxx.lanl.gov/abs/hep-th/9308133}{{\tt hep-th/9308133}}.

\bibitem{Hull:2004in}
C.~Hull, {\it {A Geometry for non-geometric string backgrounds}},  {\em JHEP}
  {\bf 0510} (2005) 065, [\href{http://xxx.lanl.gov/abs/hep-th/0406102}{{\tt
  hep-th/0406102}}].

\bibitem{Hull:2009mi}
C.~Hull and B.~Zwiebach, {\it {Double Field Theory}},  {\em JHEP} {\bf 0909}
  (2009) 099, [\href{http://xxx.lanl.gov/abs/0904.4664}{{\tt
  arXiv:0904.4664}}].

\bibitem{Hohm:2010jy}
O.~Hohm, C.~Hull, and B.~Zwiebach, {\it {Background Independent Action for
  Double Field Theory}},  {\em JHEP} {\bf 1007} (2010) 016,
  [\href{http://xxx.lanl.gov/abs/1003.5027}{{\tt arXiv:1003.5027}}].

\bibitem{Hohm:2010pp}
O.~Hohm, C.~Hull, and B.~Zwiebach, {\it {Generalized Metric Formulation of
  Double Field Theory}},  {\em JHEP} {\bf 1008} (2010) 008,
  [\href{http://xxx.lanl.gov/abs/1006.4823}{{\tt arXiv:1006.4823}}].

\bibitem{Hohm:2010xe}
O.~Hohm and S.~K. Kwak, {\it {Frame-like Geometry of Double Field Theory}},
  {\em J.Phys.} {\bf A44} (2011) 085404,
  [\href{http://xxx.lanl.gov/abs/1011.4101}{{\tt arXiv:1011.4101}}].

\bibitem{Hohm:2011nu}
O.~Hohm and S.~K. Kwak, {\it {N=1 Supersymmetric Double Field Theory}},  {\em
  JHEP} {\bf 1203} (2012) 080, [\href{http://xxx.lanl.gov/abs/1111.7293}{{\tt
  arXiv:1111.7293}}].

\bibitem{Jeon:2010rw}
I.~Jeon, K.~Lee, and J.-H. Park, {\it {Differential Geometry with a Projection:
  Application to Double Field Theory}},  {\em JHEP} {\bf 1104} (2011) 014,
  [\href{http://xxx.lanl.gov/abs/1011.1324}{{\tt arXiv:1011.1324}}].

\bibitem{Jeon:2011cn}
I.~Jeon, K.~Lee, and J.-H. Park, {\it {Stringy Differential Geometry, Beyond
  Riemann}},  {\em Phys.Rev.} {\bf D84} (2011) 044022,
  [\href{http://xxx.lanl.gov/abs/1105.6294}{{\tt arXiv:1105.6294}}].

\bibitem{Jeon:2011vx}
I.~Jeon, K.~Lee, and J.-H. Park, {\it {Incorporation of Fermions into Double
  Field Theory}},  {\em JHEP} {\bf 1111} (2011) 025,
  [\href{http://xxx.lanl.gov/abs/1109.2035}{{\tt arXiv:1109.2035}}].

\bibitem{Jeon:2011sq}
I.~Jeon, K.~Lee, and J.-H. Park, {\it {Supersymmetric Double Field Theory:
  Stringy Reformulation of Supergravity}},  {\em Phys.Rev.} {\bf D85} (2012)
  081501, [\href{http://xxx.lanl.gov/abs/1112.0069}{{\tt arXiv:1112.0069}}].

\bibitem{Blair:2013noa}
C.~D. Blair, E.~Malek, and A.~J. Routh, {\it {An $O(D, D)$ invariant
  Hamiltonian action for the superstring}},  {\em Class.Quant.Grav.} {\bf 31}
  (2014), no.~20 205011, [\href{http://xxx.lanl.gov/abs/1308.4829}{{\tt
  arXiv:1308.4829}}].

\bibitem{Berman:2013uda}
D.~S. Berman, C.~D. Blair, E.~Malek, and M.~J. Perry, {\it {The $O_{D,D}$
  geometry of string theory}},  {\em Int.J.Mod.Phys.} {\bf A29} (2014), no.~15
  1450080, [\href{http://xxx.lanl.gov/abs/1303.6727}{{\tt arXiv:1303.6727}}].

\bibitem{Blair:2014kla}
C.~D.~A. Blair, {\it {Non-commutativity and non-associativity of the doubled
  string in non-geometric backgrounds}},
  \href{http://xxx.lanl.gov/abs/1405.2283}{{\tt arXiv:1405.2283}}.

\bibitem{Wu:2013sha}
H.~Wu and H.~Yang, {\it {Double Field Theory Inspired Cosmology}},  {\em JCAP}
  {\bf 1407} (2014) 024, [\href{http://xxx.lanl.gov/abs/1307.0159}{{\tt
  arXiv:1307.0159}}].

\bibitem{Wu:2013ixa}
H.~Wu and H.~Yang, {\it {New Cosmological Signatures from Double Field
  Theory}},  \href{http://xxx.lanl.gov/abs/1312.5580}{{\tt arXiv:1312.5580}}.

\bibitem{Ma:2014ala}
C.-T. Ma and C.-M. Shen, {\it {Cosmological Implications from O(D,D)}},  {\em
  Fortsch.Phys.} {\bf 62} (2014) 921--941,
  [\href{http://xxx.lanl.gov/abs/1405.4073}{{\tt arXiv:1405.4073}}].

\bibitem{Hull:2007zu}
C.~Hull, {\it {Generalised Geometry for M-Theory}},  {\em JHEP} {\bf 0707}
  (2007) 079, [\href{http://xxx.lanl.gov/abs/hep-th/0701203}{{\tt
  hep-th/0701203}}].

\bibitem{Pacheco:2008ps}
P.~P. Pacheco and D.~Waldram, {\it {M-theory, exceptional generalised geometry
  and superpotentials}},  {\em JHEP} {\bf 0809} (2008) 123,
  [\href{http://xxx.lanl.gov/abs/0804.1362}{{\tt arXiv:0804.1362}}].

\bibitem{Hillmann:2009ci}
C.~Hillmann, {\it {Generalized E(7(7)) coset dynamics and D=11 supergravity}},
  {\em JHEP} {\bf 0903} (2009) 135,
  [\href{http://xxx.lanl.gov/abs/0901.1581}{{\tt arXiv:0901.1581}}].

\bibitem{Berman:2010is}
D.~S. Berman and M.~J. Perry, {\it {Generalized Geometry and M theory}},  {\em
  JHEP} {\bf 1106} (2011) 074, [\href{http://xxx.lanl.gov/abs/1008.1763}{{\tt
  arXiv:1008.1763}}].

\bibitem{Coimbra:2011ky}
A.~Coimbra, C.~Strickland-Constable, and D.~Waldram, {\it {$E_{d(d)} \times
  \mathbb{R}^+$ generalised geometry, connections and M theory}},  {\em JHEP}
  {\bf 1402} (2014) 054, [\href{http://xxx.lanl.gov/abs/1112.3989}{{\tt
  arXiv:1112.3989}}].

\bibitem{Coimbra:2012af}
A.~Coimbra, C.~Strickland-Constable, and D.~Waldram, {\it {Supergravity as
  Generalised Geometry II: $E_{d(d)}\times \mathbb{R}^+$ and M theory}},  {\em
  JHEP} {\bf 1403} (2014) 019, [\href{http://xxx.lanl.gov/abs/1212.1586}{{\tt
  arXiv:1212.1586}}].

\bibitem{Berman:2011pe}
D.~S. Berman, H.~Godazgar, and M.~J. Perry, {\it {SO(5,5) duality in M-theory
  and generalized geometry}},  {\em Phys.Lett.} {\bf B700} (2011) 65--67,
  [\href{http://xxx.lanl.gov/abs/1103.5733}{{\tt arXiv:1103.5733}}].

\bibitem{Berman:2011kg}
D.~S. Berman, E.~T. Musaev, and M.~J. Perry, {\it {Boundary Terms in
  Generalized Geometry and doubled field theory}},  {\em Phys.Lett.} {\bf B706}
  (2011) 228--231, [\href{http://xxx.lanl.gov/abs/1110.3097}{{\tt
  arXiv:1110.3097}}].

\bibitem{Berman:2011cg}
D.~S. Berman, H.~Godazgar, M.~Godazgar, and M.~J. Perry, {\it {The Local
  symmetries of M-theory and their formulation in generalised geometry}},  {\em
  JHEP} {\bf 1201} (2012) 012, [\href{http://xxx.lanl.gov/abs/1110.3930}{{\tt
  arXiv:1110.3930}}].

\bibitem{Berman:2011jh}
D.~S. Berman, H.~Godazgar, M.~J. Perry, and P.~West, {\it {Duality Invariant
  Actions and Generalised Geometry}},  {\em JHEP} {\bf 1202} (2012) 108,
  [\href{http://xxx.lanl.gov/abs/1111.0459}{{\tt arXiv:1111.0459}}].

\bibitem{Berman:2012vc}
D.~S. Berman, M.~Cederwall, A.~Kleinschmidt, and D.~C. Thompson, {\it {The
  Gauge Structure of Generalised Diffeomorphisms}},  {\em JHEP} {\bf 1301}
  (2013) 064, [\href{http://xxx.lanl.gov/abs/1208.5884}{{\tt
  arXiv:1208.5884}}].

\bibitem{Park:2013gaj}
J.-H. Park and Y.~Suh, {\it {U-geometry : SL(5)}},  {\em JHEP} {\bf 1304}
  (2013) 147, [\href{http://xxx.lanl.gov/abs/1302.1652}{{\tt
  arXiv:1302.1652}}].

\bibitem{Cederwall:2013oaa}
M.~Cederwall, {\it {Non-gravitational exceptional supermultiplets}},  {\em
  JHEP} {\bf 1307} (2013) 025, [\href{http://xxx.lanl.gov/abs/1302.6737}{{\tt
  arXiv:1302.6737}}].

\bibitem{Cederwall:2013naa}
M.~Cederwall, J.~Edlund, and A.~Karlsson, {\it {Exceptional geometry and tensor
  fields}},  {\em JHEP} {\bf 1307} (2013) 028,
  [\href{http://xxx.lanl.gov/abs/1302.6736}{{\tt arXiv:1302.6736}}].

\bibitem{Strickland-Constable:2013xta}
C.~Strickland-Constable, {\it {Subsectors, Dynkin Diagrams and New Generalised
  Geometries}},  \href{http://xxx.lanl.gov/abs/1310.4196}{{\tt
  arXiv:1310.4196}}.

\bibitem{Park:2014una}
J.-H. Park and Y.~Suh, {\it {U-gravity: SL(N)}},  {\em JHEP} {\bf 1406} (2014)
  102, [\href{http://xxx.lanl.gov/abs/1402.5027}{{\tt arXiv:1402.5027}}].

\bibitem{West:2001as}
P.~C. West, {\it {E(11) and M theory}},  {\em Class.Quant.Grav.} {\bf 18}
  (2001) 4443--4460, [\href{http://xxx.lanl.gov/abs/hep-th/0104081}{{\tt
  hep-th/0104081}}].

\bibitem{Englert:2003zs}
F.~Englert, L.~Houart, A.~Taormina, and P.~C. West, {\it {The Symmetry of M
  theories}},  {\em JHEP} {\bf 0309} (2003) 020,
  [\href{http://xxx.lanl.gov/abs/hep-th/0304206}{{\tt hep-th/0304206}}].

\bibitem{West:2003fc}
P.~C. West, {\it {E(11), SL(32) and central charges}},  {\em Phys.Lett.} {\bf
  B575} (2003) 333--342, [\href{http://xxx.lanl.gov/abs/hep-th/0307098}{{\tt
  hep-th/0307098}}].

\bibitem{Kleinschmidt:2003jf}
A.~Kleinschmidt and P.~C. West, {\it {Representations of G+++ and the role of
  space-time}},  {\em JHEP} {\bf 0402} (2004) 033,
  [\href{http://xxx.lanl.gov/abs/hep-th/0312247}{{\tt hep-th/0312247}}].

\bibitem{West:2004kb}
P.~C. West, {\it {E(11) origin of brane charges and U-duality multiplets}},
  {\em JHEP} {\bf 0408} (2004) 052,
  [\href{http://xxx.lanl.gov/abs/hep-th/0406150}{{\tt hep-th/0406150}}].

\bibitem{Berman:2013eva}
D.~S. Berman and D.~C. Thompson, {\it {Duality Symmetric String and M-Theory}},
   {\em Phys.Rept.} {\bf 566} (2014) 1--60,
  [\href{http://xxx.lanl.gov/abs/1306.2643}{{\tt arXiv:1306.2643}}].

\bibitem{Hohm:2013bwa}
O.~Hohm, D.~Lust, and B.~Zwiebach, {\it {The Spacetime of Double Field Theory:
  Review, Remarks, and Outlook}},  {\em Fortsch.Phys.} {\bf 61} (2013)
  926--966, [\href{http://xxx.lanl.gov/abs/1309.2977}{{\tt arXiv:1309.2977}}].

\bibitem{Aldazabal:2013sca}
G.~Aldazabal, D.~Marques, and C.~Nunez, {\it {Double Field Theory: A
  Pedagogical Review}},  {\em Class.Quant.Grav.} {\bf 30} (2013) 163001,
  [\href{http://xxx.lanl.gov/abs/1305.1907}{{\tt arXiv:1305.1907}}].

\bibitem{Sorkin:1983ns}
R.~D. Sorkin, {\it {Kaluza-Klein Monopole}},  {\em Phys.Rev.Lett.} {\bf 51}
  (1983) 87--90.

\bibitem{Gross:1983hb}
D.~J. Gross and M.~J. Perry, {\it {Magnetic Monopoles in Kaluza-Klein
  Theories}},  {\em Nucl.Phys.} {\bf B226} (1983) 29--48.

\bibitem{Townsend:1995kk}
P.~Townsend, {\it {The eleven-dimensional supermembrane revisited}},  {\em
  Phys.Lett.} {\bf B350} (1995) 184--187,
  [\href{http://xxx.lanl.gov/abs/hep-th/9501068}{{\tt hep-th/9501068}}].

\bibitem{Berkeley:2014nza}
J.~Berkeley, D.~S. Berman, and F.~J. Rudolph, {\it {Strings and Branes are
  Waves}},  {\em JHEP} {\bf 1406} (2014) 006,
  [\href{http://xxx.lanl.gov/abs/1403.7198}{{\tt arXiv:1403.7198}}].

\bibitem{Jensen:2011jna}
S.~Jensen, {\it {The KK-Monopole/NS5-Brane in Doubled Geometry}},  {\em JHEP}
  {\bf 1107} (2011) 088, [\href{http://xxx.lanl.gov/abs/1106.1174}{{\tt
  arXiv:1106.1174}}].

\bibitem{Kimura:2013fda}
T.~Kimura and S.~Sasaki, {\it {Gauged Linear Sigma Model for Exotic
  Five-brane}},  {\em Nucl.Phys.} {\bf B876} (2013) 493--508,
  [\href{http://xxx.lanl.gov/abs/1304.4061}{{\tt arXiv:1304.4061}}].

\bibitem{Kimura:2013zva}
T.~Kimura and S.~Sasaki, {\it {Worldsheet instanton corrections to
  $5^2_2$-brane geometry}},  {\em JHEP} {\bf 1308} (2013) 126,
  [\href{http://xxx.lanl.gov/abs/1305.4439}{{\tt arXiv:1305.4439}}].

\bibitem{Kimura:2013khz}
T.~Kimura and S.~Sasaki, {\it {Worldsheet Description of Exotic Five-brane with
  Two Gauged Isometries}},  {\em JHEP} {\bf 1403} (2014) 128,
  [\href{http://xxx.lanl.gov/abs/1310.6163}{{\tt arXiv:1310.6163}}].

\bibitem{Kimura:2014upa}
T.~Kimura, S.~Sasaki, and M.~Yata, {\it {World-volume Effective Actions of
  Exotic Five-branes}},  {\em JHEP} {\bf 1407} (2014) 127,
  [\href{http://xxx.lanl.gov/abs/1404.5442}{{\tt arXiv:1404.5442}}].

\bibitem{Verlinde:1995mz}
E.~P. Verlinde, {\it {Global aspects of electric - magnetic duality}},  {\em
  Nucl.Phys.} {\bf B455} (1995) 211--228,
  [\href{http://xxx.lanl.gov/abs/hep-th/9506011}{{\tt hep-th/9506011}}].

\bibitem{Witten:1995gf}
E.~Witten, {\it {On S duality in Abelian gauge theory}},  {\em Selecta Math.}
  {\bf 1} (1995) 383, [\href{http://xxx.lanl.gov/abs/hep-th/9505186}{{\tt
  hep-th/9505186}}].

\bibitem{Ortin04}
T.~Ort\'{i}n, {\em {Gravity and Strings}}.
\newblock Cambridge University Press, 2004.

\bibitem{Papadopoulos:2014mxa}
G.~Papadopoulos, {\it {Seeking the balance: Patching double and exceptional
  field theories}},  {\em JHEP} {\bf 1410} (2014) 89,
  [\href{http://xxx.lanl.gov/abs/1402.2586}{{\tt arXiv:1402.2586}}].

\bibitem{Berman:2014jba}
D.~S. Berman, M.~Cederwall, and M.~J. Perry, {\it {Global aspects of double
  geometry}},  {\em JHEP} {\bf 1409} (2014) 066,
  [\href{http://xxx.lanl.gov/abs/1401.1311}{{\tt arXiv:1401.1311}}].

\bibitem{Cederwall:2014opa}
M.~Cederwall, {\it {T-duality and non-geometric solutions from double
  geometry}},  {\em Fortsch.Phys.} {\bf 62} (2014) 942--949,
  [\href{http://xxx.lanl.gov/abs/1409.4463}{{\tt arXiv:1409.4463}}].

\bibitem{Tong:2002rq}
D.~Tong, {\it {NS5-branes, T duality and world sheet instantons}},  {\em JHEP}
  {\bf 0207} (2002) 013, [\href{http://xxx.lanl.gov/abs/hep-th/0204186}{{\tt
  hep-th/0204186}}].

\bibitem{Harvey:2005ab}
J.~A. Harvey and S.~Jensen, {\it {Worldsheet instanton corrections to the
  Kaluza-Klein monopole}},  {\em JHEP} {\bf 0510} (2005) 028,
  [\href{http://xxx.lanl.gov/abs/hep-th/0507204}{{\tt hep-th/0507204}}].

\bibitem{Kimura:2014bxa}
T.~Kimura and M.~Yata, {\it {Gauged Linear Sigma Model with F-term for A-type
  ALE Space}},  {\em PTEP} {\bf 2014} (2014), no.~7 073B01,
  [\href{http://xxx.lanl.gov/abs/1402.5580}{{\tt arXiv:1402.5580}}].

\bibitem{Kimura:2014aja}
T.~Kimura and M.~Yata, {\it {T-duality Transformation of Gauged Linear Sigma
  Model with F-term}},  {\em Nucl.Phys.} {\bf B887} (2014) 136--167,
  [\href{http://xxx.lanl.gov/abs/1406.0087}{{\tt arXiv:1406.0087}}].

\bibitem{deBoer:2010ud}
J.~de~Boer and M.~Shigemori, {\it {Exotic branes and non-geometric
  backgrounds}},  {\em Phys.Rev.Lett.} {\bf 104} (2010) 251603,
  [\href{http://xxx.lanl.gov/abs/1004.2521}{{\tt arXiv:1004.2521}}].

\bibitem{deBoer:2012ma}
J.~de~Boer and M.~Shigemori, {\it {Exotic Branes in String Theory}},  {\em
  Phys.Rept.} {\bf 532} (2013) 65--118,
  [\href{http://xxx.lanl.gov/abs/1209.6056}{{\tt arXiv:1209.6056}}].

\bibitem{Hohm:2013pua}
O.~Hohm and H.~Samtleben, {\it {Exceptional Form of D=11 Supergravity}},  {\em
  Phys.Rev.Lett.} {\bf 111} (2013) 231601,
  [\href{http://xxx.lanl.gov/abs/1308.1673}{{\tt arXiv:1308.1673}}].

\bibitem{Hohm:2013vpa}
O.~Hohm and H.~Samtleben, {\it {Exceptional Field Theory I: $E_{6(6)}$
  covariant Form of M-Theory and Type IIB}},  {\em Phys.Rev.} {\bf D89} (2014)
  066016, [\href{http://xxx.lanl.gov/abs/1312.0614}{{\tt arXiv:1312.0614}}].

\bibitem{Hohm:2013uia}
O.~Hohm and H.~Samtleben, {\it {Exceptional Field Theory II: E$_{7(7)}$}},
  {\em Phys.Rev.} {\bf D89} (2014) 066017,
  [\href{http://xxx.lanl.gov/abs/1312.4542}{{\tt arXiv:1312.4542}}].

\bibitem{Hohm:2014fxa}
O.~Hohm and H.~Samtleben, {\it {Exceptional Field Theory III: E$_{8(8)}$}},
  {\em Phys.Rev.} {\bf D90} (2014) 066002,
  [\href{http://xxx.lanl.gov/abs/1406.3348}{{\tt arXiv:1406.3348}}].

\bibitem{Godazgar:2014nqa}
H.~Godazgar, M.~Godazgar, O.~Hohm, H.~Nicolai, and H.~Samtleben, {\it
  {Supersymmetric E$_{7(7)}$ Exceptional Field Theory}},  {\em JHEP} {\bf 1409}
  (2014) 044, [\href{http://xxx.lanl.gov/abs/1406.3235}{{\tt
  arXiv:1406.3235}}].

\bibitem{Berman:2014hna}
D.~S. Berman and F.~J. Rudolph, {\it {Strings, Branes and the Self-dual
  Solutions of Exceptional Field Theory}},  {\em JHEP} {\bf 1505} (2015) 130,
  [\href{http://xxx.lanl.gov/abs/1412.2768}{{\tt arXiv:1412.2768}}].

\bibitem{Berman:2007tf}
D.~S. Berman and L.~C. Tadrowski, {\it {M-theory brane deformations}},  {\em
  Nucl.Phys.} {\bf B795} (2008) 201--229,
  [\href{http://xxx.lanl.gov/abs/0709.3059}{{\tt arXiv:0709.3059}}].

\bibitem{Aharony:1998ub}
O.~Aharony, M.~Berkooz, D.~Kutasov, and N.~Seiberg, {\it {Linear dilatons, NS
  five-branes and holography}},  {\em JHEP} {\bf 9810} (1998) 004,
  [\href{http://xxx.lanl.gov/abs/hep-th/9808149}{{\tt hep-th/9808149}}].

\end{thebibliography}\endgroup

\end{document}